\documentclass[paper,12pt]{JHEP}
\usepackage[centertags]{amsmath}
\usepackage{amsfonts} \usepackage{amssymb} \usepackage{amsthm}
\allowdisplaybreaks[1]
\usepackage{graphicx}

\newcommand{\R}{{\mathbb R}}
\newcommand{\C}{{\mathbb C}}

\newcommand{\ii}{\mathrm{i}}
\newcommand{\ee}{\mathrm{e}}

\newcommand{\one}{{\rm 1\kern -.9mm l}}

\newcommand{\ft}[2]{{\textstyle\frac{#1}{#2}}}

\newdimen\tableauside\tableauside=1.0ex
\newdimen\tableaurule\tableaurule=0.4pt
\newdimen\tableaustep
\def\phantomhrule#1{\hbox{\vbox to0pt{\hrule height\tableaurule
width#1\vss}}}
\def\phantomvrule#1{\vbox{\hbox to0pt{\vrule width\tableaurule
height#1\hss}}}
\def\sqr{\vbox{%
  \phantomhrule\tableaustep
\hbox{\phantomvrule\tableaustep\kern\tableaustep\phantomvrule\tableaustep}%
  \hbox{\vbox{\phantomhrule\tableauside}\kern-\tableaurule}}}
\def\squares#1{\hbox{\count0=#1\noindent\loop\sqr
  \advance\count0 by-1 \ifnum\count0>0\repeat}}
\def\tableau#1{\vcenter{\offinterlineskip
  \tableaustep=\tableauside\advance\tableaustep by-\tableaurule
  \kern\normallineskip\hbox
    {\kern\normallineskip\vbox
      {\gettableau#1 0 }%
     \kern\normallineskip\kern\tableaurule}%
  \kern\normallineskip\kern\tableaurule}}
\def\gettableau#1 {\ifnum#1=0\let\next=\null\else
  \squares{#1}\let\next=\gettableau\fi\next}
\def\XXint#1#2#3{{\setbox0=\hbox{$#1{#2#3}{\int}$}
     \vcenter{\hbox{$#2#3$}}\kern-.5\wd0}}

\def\be{\begin{equation}}
\def\ee{\end{equation}}
\def\bea{\begin{eqnarray}}
\def\eea{\end{eqnarray}}
\newcommand{\nn}{\nonumber}
\def\ii{{\rm i}}

\title{Stringy origin of 4d black hole microstates 
}
\author{ M. Bianchi$^{1}$,  J.F. Morales$^{2}$ and L. Pieri$^{1}$
\\
\vskip 0.2cm
$^1$ Universit\`a di Roma Tor Vergata, Dipartimento di Fisica
\\ Via della Ricerca Scientifica, I-00133 Roma, Italy\\
\vskip 0.2cm
$^2$ I.N.F.N - sezione di Roma 2\\
and Universit\`a di Roma Tor Vergata, Dipartimento di Fisica\\
Via della Ricerca Scientifica, I-00133 Roma, Italy
\vspace{0.35cm}
\email{bianchi,morales,lorenzo.pieri@roma2.infn.it}
}
\abstract{ We  derive a precise dictionary between  micro-state geometries and  open string condensates for a large class of excitations of four dimensional  BPS black holes realised in terms of D3-branes intersecting on a six-torus.  The complete multipole expansion of the supergravity solutions at weak coupling is extracted  from string amplitudes involving one massless closed string and multiple open strings insertions on disks with mixed boundary conditions. }


\preprint{ROM2F/2016/04}

\begin{document}

\date

\tableofcontents

\pagebreak

\section{Introduction}

The fuzzball proposal \cite{Mathur:2009hf,Mathur:2005zp} relates black-hole micro-states to smooth and horizon-less geometries  with the same asymptotics (the same mass and charges) as the putative black hole but  differing from it in its interior. The (semi)classical black-hole geometry only results from a coarse-graining approximation of the micro-state solutions.
For black holes corresponding to D-brane bound-states, the micro-states can be realised and counted in terms of open-string excitations of the D-branes constituents \cite{Sen:1995in, Strominger:1996sh}. As a result, the fuzzball proposal suggests a  one-to-one correspondence between open-string condensates and the geometries generated by them. 
 
 The best understood example is the two-charge BPS system in type IIB theory for which the whole family of micro-state geometries is known \cite{Lunin:2001fv,Lunin:2002bj,Kanitscheider:2007wq}.
 The solutions can be written  in several different U-duality related frames.  In  \cite{Giusto:2009qq}, the asymptotic form of the general micro-state geometries associated to a system of D1- and D5-branes  has been derived from disk amplitudes computing the emission rate of closed  string states from open string condensates at the D1-D5 intersections. The two-charge system represents a somewhat degenerate example of a black hole in that its classical geometry has a singular horizon of zero size. Smooth horizons in the supergravity limit require systems with 3- or 4-charges. Unfortunately, despite significant progress (see \cite{Bena:2007kg}\nocite{Skenderis:2008qn,Mathur:2008nj, Giusto:2011fy, Giusto:2012yz, Bena:2011uw, Giusto:2012jx, Giusto:2013bda,Bena:2014qxa}-\cite{Bena:2015bea} and references therein for  recent  results and reviews),  a similar analysis for 3- and 4-charge black holes is still missing since a family of micro-state solutions large enough to account for the black hole entropy is not yet known. 
 
 Aim of this paper is to partially fill in this gap, providing a direct link between micro-state geometries and open string condensates for a large class  of 4-charge black hole microstates. We realise the four-dimensional BPS black holes in the `democratic' frame of  four mutually intersecting stacks of D3-branes on  $T^6$. We focus on a family of black hole solutions characterised by eight (`lovely' rather than `hateful') harmonic functions in the type IIB theory.  This family is dual to the extensively studied BPS solutions \cite{Denef:2000nb}\nocite{Denef:2002ru,Bates:2003vx,Balasubramanian:2006gi,deBoer:2008fk}-\cite{Dall'Agata:2010dy} describing  D0-D2-D4-D6 brane bound-states in the type IIA theory or 
   intersecting M2 and M5 branes in M-theory carrying non-trivial momentum and KK monopole charge. Amusingly,  in the D3-brane frame, the eight harmonic functions are described entirely in terms of excitations of the metric and the four-form RR field. A subclass of this family of solutions involving two harmonic functions has been studied recently in \cite{Lunin:2015hma} and shown to be regular and horizon-less.
 
 The micro-state geometries will be derived from disk amplitudes  computing the  emission rate of massless closed string states from open string condensates  binding the different stacks of D3 branes. The disk amplitude computes the  leading correction to the supergravity solution  in the limit of small string coupling  $g_s$. In principle the supergravity solution can be  recovered by plugging the result  into the type IIB equations of motion and solving recursively order by order in $g_s$. Equivalently, one can expand the general micro-state  solution to order $g_s$ and match it against the result of the string amplitude.  Proceeding in this way, we identify three class of micro-states depending on whether the open string condensate involves either one D3-brane stack or the intersection of two or four different stacks. We dub the three classes as 
 L, K and M solutions referring to the names of the harmonic functions underlying the solutions. We remark that disk contributions add  and at this linear order both the metric and the four-form potential can be written as a sum of harmonic functions. In particular, we find that four of the  harmonic functions (L) specifying the solution are associated to the four different disks  with a single boundary, three harmonic functions  (K) are associated to the three inequivalent 2-charge intersections and the last one (M) is sourced by a 4-charge open string condensate. 
 The three types of micro-state geometries fall off at infinity as $Q_i/r$, $Q_i Q_j/r^2$ and $Q_1 Q_2 Q_3 Q_4/r^3$ respectively.  Higher multipole modes are generated by insertions of untwisted (boundary-preserving) open strings on the boundary of the disk, much in the same way as for the D1-D5 system \cite{Giusto:2011fy}. Remarkably, as we will see, the micro-state angular momentum is fully accounted for by the type K component of the solution.  
 
Solutions of type M  are particularly interesting  since the  number of disks with four different boundaries grows with the product of the four charges $N=Q_1\, Q_2\, Q_3 \,Q_4$,  the same quartic invariant  entering  in the   black-hole entropy formula $S= 2\pi \sqrt{ N }$.  It is tempting to speculate that after quantization the corresponding micro-states may account for the entropy of the putative black-hole  much in the same way as the entropy  of the 2-charge system with $N=Q_1 Q_2$ is reproduced by quantisation of the profile function describing D1-D5 intersections \cite{Lunin:2001fv,Rychkov:2005ji,Krishnan:2015vha}  or, equivalently, D3-D3' pair intersections, in the systems that we consider here\footnote{Indeed, in our case, any pair of D3-branes intersect along a string in $T^6$ and can be mapped into D1-D5 or F1-P1 via U-dualities.}.

We remark that the class of 4-charge micro-state solutions generated by  open string condensates that we find here is larger than the family of supergravity solutions explicitly known. This is at variant with the 2-charge systems, where the most general micro-state \cite{Kanitscheider:2007wq} is reproduced by a 2-charge intersection \cite{Giusto:2009qq}. In the case of 4 charges we find new solutions looking pretty  much like the K and M solutions, but tilted  with respect to the background D3-brane geometry.  It would be interesting to understand whether these could be accommodated in a larger class of micro-state geometries or should be excluded on the basis of supersymmetry or stability conditions (D- and F-flatness) for the open strings.

The plan of the paper is as follows. In Section 1 we  introduce the `eight-ful'  harmonic family of supergravity solutions in the  `democratic' frame of  D3-branes. We identify three sub-classes of solutions that we dub L solutions, K solutions and M solutions. 
In Section 2, we focus on solutions sourced by disks with a single type of boundary condition.  
We compute disk amplitudes involving the insertion of a closed string NS-NS or R-R vertex operator   in the bulk and an arbitrary number of untwisted open string
scalars.
In Section 3, we compute  disk amplitudes with two twisted open string insertions that intertwine between two different boundary conditions. 
In Section 4, we compute disk amplitudes with four different boundaries conditions. The computations are much more involved (they are equivalent to a 6-point open string amplitude) but one can numerically integrate the final correlators and get a finite result both for NS-NS and R-R closed string insertions in the bulk. Remarkably, NS-NS and R-R correlators are proportional to the same integrals, so their numeric values are irrelevant for the  identification of the dual supergravity solution.
We finally draw our conclusions and indicate directions for further research. 
In the appendices we collect our conventions for string correlators and  present the derivation of the D3-brane solution starting from its T-dual description  in Type IIA theory.   

 \section{Supergravity solution}
 
 We consider  four-dimensional BPS black hole solutions of type IIB supergravity  consisting of D3-branes intersecting on $T^6$.    
 The solutions  are characterised by a non-trivial metric and a five-form flux with all the other fields of type IIB supergravity vanishing. 
 In this sector, the type IIB equations of motion take the extremely simple and elegant form\footnote{Pretty much as for gauge fields in $D=4$, the stress tensor of a 4-form is (classically) trace-less in $D=10$.}
 \bea
 R_{MN} &=& {1\over 4\cdot 4!} F_{M P_1 P_2 P_3 P_4} \, F_{N}{}^{P_1 P_2 P_3 P_4}\nn\\
 F_5 &=&  {*_{10}}F_5
 \eea
 with
 \be
 F_5=d C_4
 \ee
   A family of solutions  can be found by acting with three T-dualities on the more familiar class \cite{Balasubramanian:2006gi,Dall'Agata:2010dy}  describing
   systems of D0-D2-D4-D6 branes in type IIA theory   (see  Appendix  for details). 
    The solutions can be written in  terms of eight harmonic functions 
  ($a=1,\dots 8$, $I=1,2,3$) 
   \be
   H_a=\{ V, L_I, K_I, M \} \label{harmh}
   \ee   
  on (flat) $\mathbb{R}^3$ associated to the four electric and four magnetic charges in the type IIA description of the system. 
   These functions are conveniently combined into  
  \bea
 P_I &=& {K_I \over V} \nn\\
 Z_I &=& L_I +{ |\epsilon_{IJK}|\over 2} {K_J K_K\over V} \nn\\
 \mu &=& { M\over 2} +{L_I K_I \over 2\, V}+{ |\epsilon_{IJK}| \over 6} \, {K_I K_J K_K \over V^2}  \label{mudef}
 \eea
 Here
  $\epsilon_{IJK}$ characterise the triple  intersections  between two cycles on $T^6$.
  
In terms of these functions the type IIB supergravity solution can be written as
      \bea
  ds^2  &=& -   e^{2U}( dt+w)^2 +e^{-2U} \,  \sum_{i=1}^3 dx_i^2 +   \sum_{I=1}^3   \left[  { d y_I^2 \over  V e^{2U} Z_I }  + V e^{2U} Z_I  \,  \tilde e_I^2  \right] \nn\\
  C_4 &=& \alpha_0  \wedge \tilde e_1\wedge \tilde e_2\wedge \tilde e_3+ \beta_0  \wedge dy_1\wedge  dy_2\wedge  dy_3 \nn\\ 
  && +\ft12 \epsilon_{IJK}\,   \left( \alpha_I  \wedge  dy_I \wedge \tilde e_J   \wedge \tilde e_K + \beta_I  \wedge  \tilde e_I \wedge dy_J   \wedge dy_K  \right)   \label{d34}
 \eea
 with  
  \bea
  e^{-4U} &=&  Z_1 Z_2 Z_3 V-\mu^2  V^2 \qquad ~~~~~~~~~~ \nn \\
   b_I &=&     P_I-\frac{\mu}{Z_I}    \qquad \qquad ~~~~~~~~~~~~~~~~~ \tilde e_I =   d \tilde y_I   -  b_I\, dy_I   \nn\\
 \alpha_0 &=& A-\mu\, V^2\, e^{4U}\, (dt+w) \qquad ~~~~~
\alpha_I =   -\frac{(dt+w)}{Z_I} + b_I\, A+w_I   \nn\\
\beta_0&=& -v_0 +{e^{-4U}\over V^2 Z_1 Z_2 Z_3}(dt+w)-b_I \,v_I+ b_1 \,b_2\, b_3 \, A +{|\epsilon_{IJK}|\over 2}\,b_I \,b_J\, w_K  \nn\\
\beta_I &=& -v_I + {|\epsilon_{IJK}|\over 2}\, \left(  {\mu\over Z_J Z_K}\, (dt+w)+b_J \, b_K\, A+2 b_J\, w_K  \right) 
 \eea
   and
  \bea
 {*_3}dA &=& d V     \qquad {*_3}dw_I = -d (K_I)  \qquad    {*_3}dv_0 =dM  \qquad    {*_3}dv_I =dL_I\nn\\
   \qquad  {*_3}dw  &=&  V d \mu-\mu dV-V Z_I dP_I  
  \eea
        
  For $K_I=M=0$,   the supergravity solution (\ref{d34})  describes a system of four stacks of intersecting D3-branes aligned according to     
   the table below. 
   The four harmonic functions $V,L_I$ describe the distribution of the D3-branes in the non compact spatial directions $x\in \mathbb{R}^3$.    
    
 \begin{table}[h]
\begin{center}
\begin{tabular}{|c|c|c|c|c|c|c|c|c|c|c|}
\hline
        Brane     &  t       & $x_1$   &  $x_2$  & $x_3$   & $ y_1 $ & $ \tilde y_1$    & $ y_2 $ & $ \tilde y_2$ & $ y_3 $ & $ \tilde y_3$        \\
\hline
 $ D3_0 $       &  $-$      &$.$    & $.$   &$.$    & $-$  &$.$   &$-$    &$.$     & $-$   & $.$     \\
$ D3_1$           &  $-$      &$.$    & $.$   &$.$    & $-$  &$.$   &$.$    &$-$     & $.$   & $-$     \\
$ D3_2$           &  $-$      &$.$    & $.$   &$.$    & $.$  &$-$   &$-$    &$.$     & $.$   & $-$     \\
 $ D3_3 $       &  $-$      &$.$    & $.$   &$.$    & $.$  &$-$   &$.$    &$-$     & $-$   & $.$     \\
\hline
\end{tabular}
\label{td3s}
\end{center}
\caption{D3-brane configuration: Neumann (N) and Dirichlet (D) directions are represented by lines and dots respectively.}
 \end{table}

  Solutions with non-zero $K_I$ and $M$ will be associated to condensates involving twisted open strings. These excitations modify the geometry of the putative black hole in its interior while preserving the same asymptotic fall 
off at infinity. Consequently, $K_I$ and $M$  should decay  faster than $|x|^{-1}$ at infinity. 
We thus focus on solutions involving harmonic functions with the following asymptotics     
 \bea
 L_I &\approx& 1+\alpha_{D3} \, {N_I\over |x| } +\ldots   \qquad     V \approx 1+ \alpha_{D3} \,{N_{0}\over |x| } +\ldots     \nn\\
 K_I & \approx&  \alpha_{D3} \, c^{K_I}_i {x^i\over |x|^3} +\ldots   \qquad  ~~  M \approx  \alpha_{D3}  \, c^M_i {x^i\over |x|^3}  +\ldots
 \label{KLMasympt}
 \eea 
 with  $c_i$ some constants, $\alpha_{D3}=4\,\pi \,g_s \, {(\alpha')^2}/{V_{T_3}} $ the inverse D3-brane tension and $V_{T_3}$  the volume of the 3-torus wrapped by the stack of D3-branes, that  for simplicity we will take  to be equal for all the stacks. We anticipate that the explicit string computations show that the coefficients $c_i$'s entering in (\ref{KLMasympt}) 
 satisfy the relation
 \be
c^M_i+\sum_{I=1}^3 c^{K_I}_i=0
 \ee
 or equivalently the function $\mu$ defined in (\ref{mudef}) starts at order $1/r^3$. 
    Since at linear order in $\alpha_{D3}$, the harmonic functions entering in the solution simply add, we can consider the contribution of each harmonic function separately.  More precisely, we define three basic class of solutions: L solutions are associated to the harmonic functions $L_I,V$ with $K_I=M=0$. K solutions are generated by turning on a single
  $K_I$ harmonic function for a given $I$ and taking $M=-K_I$  for the given I. Finally M solutions are associated to the orthogonal combination   $M+\sum_I K_I$ that is generated
  at order $1/r^3$. In other words the dipole modes in  (\ref{KLMasympt})  vanish: $c^M_i= c^{K_I}_i=0$.

 The  family (\ref{d34}) can be seen as the ``superposition" of these  three basic classes of solutions. We remark that disks emissions simply add but massless closed string states 
    interact with each other. This entails non-linearity, so the resulting solution will not be the naive  ``superposition" of its constituents as it is obvious from (\ref{d34}).  
  In the rest of this section, we  display  the supergravity solutions for the three basic classes of micro-state geometries.

 \subsection{  L solutions}   
  
  A representative of the L class of micro-state solutions is given by taking 
  \be
  V=L(x) \qquad\qquad M=K_I=0 \qquad\qquad L_I=1 
  \ee
   in (\ref{d34}). This results into
      \bea
   e^{-2U} &=& L^{1/ 2} \qquad\qquad  \tilde e_I=d\tilde y_I  \qquad\qquad \mu=0       \nn\\
   d A &=& {*_3}dL   \qquad    \alpha_0= A \qquad  \beta_0=L^{-1}\, dt  \qquad w=\beta_I=0   \qquad \alpha_I=-dt
   \eea
   The Type IIB supergravity solution reduces to  (discarding d-exact terms) 
  \bea
 ds^2 &=&  L^{-{1/2}} \left(-dt^2+ \sum_{i=1}^3 dy_i^2 \right )  +  L^{{1/2}}  \sum_{i=1}^3 \left(   dx_i^2+  d\tilde y_i^2 \right)   \nn\\
 C_4 &=&  L^{-1}  \wedge dt\wedge dy_1 \wedge  dy_2 \wedge  dy_3  +A \wedge d\tilde y_1\wedge d\tilde y_2\wedge d\tilde y_3   \label{d34l}
 \eea
  This solution is generated by distributing D3-branes of type 0 in table 1 along $\mathbb{R}^3$.
  Similar solutions are found by turning on $L_I$ associated to the
 other three types of D3-branes in the table\footnote{For a spherical symmetric   harmonic function
  $
   L(r)={Q/r}     
   $
   one finds  $A=Q\, \cos\theta \, d\varphi$    in spherical coordinates where $dx_i^2=dr^2+r^2( d\theta^2+\sin^2\theta\, d\varphi^2) $. }.

 \subsection{ K solutions}
 
  Next we consider solutions with 
  \be
  K_3 =-M=K(x)  \qquad\quad  \mu=0 \qquad\quad L_I=V=1  \qquad\quad K_1=K_2 =0 
\ee
   For this choice one finds
 \bea
   e^{-2U}&=&1\qquad {*_3}dw  = - d K \qquad \tilde e_{1}=d\tilde y_1 \qquad \tilde e_{2}=d\tilde y_2 \qquad \tilde e_3 =d\tilde y_3 -K \,dy_3 \nn   \\
    \alpha_0 &=& \beta_I=0  \qquad\quad  \alpha_1=\alpha_2=-(dt+w)  \qquad\quad \alpha_3=-dt  \qquad\quad  \beta_0=dt  
   \eea
  and the resulting   supergravity solution reduces to (discarding d-exact terms) 
\bea
 ds^2 &=& -   ( dt+w)^2 +  \sum_{i=1}^3 \left(  dx_i^2 +dy_i^2 + d\tilde y_i^2 \right)   - 2 \,K \,dy_3 \,d\tilde y_3 +K^2\, dy_3^2  \nn \\
 C_4 &=&  - (dt+w)   \wedge  \tilde e_3\wedge  (dy_1 \wedge d\tilde y_2+ d\tilde y_1 \wedge d y_2  )   
  \label{d34k}
 \eea
  Similar solutions are found by turning on $K_1$ or $K_2$.  We would like to stress that K solutions  in general carry angular momentum, unlike L and M solutions.

  \subsection{ M solutions}
 
   Finally we consider solutions with 
  \be
  K_2=M=M(x)  \qquad  \mu=M \qquad L_I=V=1  \qquad K_1=K_3 =0 
\ee
   Now one finds
 \bea
   e^{-2U}&=&\sqrt{1-M^2} \quad \tilde e_{1}=d\tilde y_1+M \, dy_1  \quad \tilde e_{2}=d\tilde y_2 \quad \tilde e_3 =d\tilde y_3 +M \,dy_3 \quad w =0    \nn\\
   \alpha_0 &=& {-M dt\over 1-M^2}\qquad\alpha_1=\alpha_3= -dt\qquad\alpha_2=-dt+w_2\qquad{dw}_2=-{*_3}dM    \\
   \beta_0 &=& (1-M^2)\,dt+(1+M^2)\, w_2 \qquad \beta_1=\beta_3=M(dt-w_2)  \qquad \beta_2=M \,dt\nn
   \eea
 The Type IIB supergravity solution reads (discarding d-exact terms) 
 \bea
 && ds^2 = -   e^{2U} dt^2 +e^{-2U} \,  \sum_{i=1}^3 (dx_i^2+dy_i^2)    +e^{2U}   \,  \left[ \sum_{i=1,3}(d \tilde y_{i}   + M \, dy_{i} )^2      + d\tilde y_2^2 \right] \label{d34mu} \\
&& C_4 =- {1\over 1-M^2} \, dt  \wedge  \tilde e_3   \wedge  ( dy_1\wedge d \tilde y_2+ M\, d\tilde y_1\wedge d  \tilde y_2 )  \nn\\
&& ~~~~~~~~~~~+ w_2 \wedge ( dy_1   \wedge dy_2 \wedge dy_3 +     d\tilde y_1 \wedge dy_2\wedge  d\tilde y_3   )  
  \eea

\subsection{Harmonic Expansions} 

The harmonic functions $H_a$  entering into the supergravity solutions   can be conveniently expanded in multi-pole modes.  The general 
expansion can be written  as
   \be
H_a(x)=h_a+\sum_{n=0}^\infty  c^a_{i_1\ldots i_n }  P_{i_1 \ldots i_n} (x)       
\label{hhh}
 \ee
  with  $ P_{i_1 \ldots i_n} (x)$ a totally symmetric and traceless rank $n$ tensor  providing a basis of harmonic functions on $\mathbb{R}^3$. The explicit
  form of  $ P_{i_1 \ldots i_n} (x) $ can be found from the multi-pole expansion
 \be
 {1\over  |x + a|} =\sum_{n=0}^\infty  a_{i_1} \ldots a_{i_n }  P_{i_1 \ldots i_n} (x)    
 \label{1multi}
 \ee  
 For example for $n=0,1,2$ one finds
 \be
 P(x)={1\over |x|} \qquad  \qquad  P_{i}(x)= - {x_i \over  |x|^3}   \qquad \qquad     P_{i j}(x)={3 x_i x_j - \delta_{ij} |x|^2 \over  |x|^5}   
 \ee
 and so on.
 We notice that even if the explicit form of $P_{i_1 \ldots i_n}(x)$  can quickly become messy as $n$ grows, the form of the
   Fourier transform is very simply. Indeed writing
  \be
   P_{i_1 \ldots i_n} (x)  = \int {d^3 k \over (2\pi)^3}   e^{i k x}  \tilde P_{i_1\ldots i_n}(k)   
  \ee
  and using
  \be
   {1\over   |x+a| }  =   \sum_{n=0}^\infty  a_{i_1} \ldots a_{i_n }  P_{i_1 \ldots i_n} (x) =4\pi   \int {d^3 k \over (2\pi)^3}  \frac{   e^{i k(x+a)}   }{k^2}  
  \ee 
one finds that the Fourier transform of the harmonic functions   $P_{i_1 \ldots i_n} (x)$ are simply polynomials of the momenta divided by $k^2$  
   \be
  \tilde P_{i_1\ldots i_n}(k)  =     { 4\pi \, (\ii)^n \over n! \, k^2}  k_{i_1} \ldots  k_{i_n}  
  \ee
   Disk amplitudes will be conveniently written in this basis.  We notice that even if the single harmonic components $P_{i_1\ldots i_n}(x)$ are singular at the origin, 
   in analogy with the 2-charge case \cite{Lunin:2002iz,Mathur:2005zp},   one may expect that for an appropriate choice of the coefficients $c_{i_1\ldots i_n}$ the infinite sum (\ref{hhh}) produce a fuzzy and smooth geometry.
This was explicitly shown in \cite{Lunin:2015hma} for a class of solutions parametrized by one complex harmonic function ${\cal H}={\cal H}_1+\ii {\cal H}_2$ obtained from our general solution by taking 

        \bea
 {\cal H}_1&=&L_1=L_2=L_3=V   \nn \\
 {\cal H}_2&=&K_1=K_2=K_3=-M  
  \eea
    
\section{ String amplitudes and 1-charge microstates }
 
 At weak coupling, the gravitational background generated by a D-brane state can be extracted from a disk amplitude  involving the insertion of a closed string states in the bulk
 and open string states specifying the D-brane micro-state on the boundary. The open string state, or condensate,  is specified by a boundary operator ${\cal O}$ given by a trace along the boundary of the disk of a product of constant open string  fields. The profile of a supergravity field $\Phi$ in the micro-state geometry can be reconstructed from the string 
 amplitude  $ {\cal A}_{\Phi,{\cal O}}(k)$  after Fourier transform in the momentum $k$ of the closed string state. Moreover, since we consider constant open string fields we can take all open string  momenta to vanish. For simplicity, we take the only non vanishing components  $k_i$ of the closed string momentum to  be along the 3 non compact space directions, {\it i.e.} $k_0=k_{y_I} = k_{\tilde y_I}=0 $. In order to keep the mass-shell condition $k^2 =0$, we analytically continue $k_i$ to complex values, and require $k_i^2 \approx 0$.
   More precisely, the deviation  $\delta\tilde \Phi(k)$ from flat space of a field $\tilde \Phi(k)$ is extracted from the string amplitude via the bulk-to-boundary  formula
      \be
 \delta\tilde \Phi(k)
  =  \left( - {\ii  \over k^2} \right) {\delta {\cal A}_{\Phi,{\cal O}}(k) \over \delta \Phi} 
  \ee
  with  $(- \ii  /k^2)$ denoting the massless closed string propagator for the (transverse) physical modes. 
  
  We notice that the limit of zero momenta for open string states 
  is well defined only in the case the disk diagram cannot factorize into two or more disks via the exchange of open string states. Indeed, in presence of factorization channels 
  the string amplitude has poles in the open string momenta and the underlying world-sheet integral diverges. We will always carefullly choose the open string  polarisations  in such a way that no factorization channels be allowed. As a result, one finds that, up to $1/k^2$, $\delta \widetilde \Phi (k)$ is a polynomial in $k_i$ and therefore can be expressed as a linear combination
  of harmonic functions  $\tilde P_{i_1\ldots i_n}(k) $ in momentum space
   \be
   \delta \widetilde \Phi (k)=  \sum_n  c^{\Phi}_{i_1\ldots i_n}  \tilde P_{i_1\ldots i_n}(k) 
   \ee
In this section we illustrate  the stringy description of the  micro-state geometry  for the very familiar case of parallel (wrapped) D3-branes. The vacua of this system are parametrised by the  expectation values of the (three) scalar fields $\phi_i$, along the non-compact directions. 
  The general open string condensate involves  an arbitrary number of insertions of the `untwisted' field $\phi_i$ on the boundary of the disk. On the other hand, the associated micro-state
  geometry is described by a D3-brane solution characterised by a smooth multi-pole harmonic function determined  by the open string condensate  
  \cite{Klebanov:1999tb}.  The string computation will allow us to fix the relative normalisation between the NS and RR vertices that will be used in the analysis of 2- and 4-charge intersections in the next sections.

 \subsection{ L solutions at weak coupling}
 
   We start by considering the supergravity solution generated by a single stack of D3-branes, let us say of type $0$. The gravity solution is given by (\ref{d34l}). 
   At linear order in $\alpha_{D3}$ one finds 
     \bea
 \delta g_{MN} dx^M dx^N  &=&   { \delta L \over 2}  \left[  dt^2- \sum_{i=1}^3 (dy_i^2 -dx_i^2-d \tilde y_i^2)    \right] +\ldots  \nn\\
  \delta C_4 &=&  -\delta L  \wedge dt\wedge dy_1 \wedge  dy_2 \wedge  dy_3  + A \wedge d\tilde y_1\wedge d\tilde y_2\wedge d\tilde y_3 +\ldots
   \label{metricv}
 \eea
     with $\delta L=L-1$ and $A$ both or order $\alpha_{D3}$.  In particular one can take
   \be
  L=1+ {\alpha_{D3} N_0\over  |x| }   + \dots  \qquad ~~~~~~  {*_3}dL=d A
    \ee
     corresponding to $N_0$ D3 branes  on $\R^3$.

%

\subsection{NS{-}NS amplitude}

 Let us start by considering a single D3-brane.
 The moduli space is parametrized in terms of  vacuum expectation value of open string scalar fields  $\phi_i$  describing the position of the brane in the transverse space. The associated supergravity solution can be extracted from disk amplitudes involving the  insertion of one closed string field and a bunch of open scalar fields $\phi_i$ at zero momenta.
 
      The open string background can be conveniently encoded in a generating function   
\be
\xi(\phi)=\sum_{n=0}^\infty \xi_{i_1\ldots i_n } \phi^{i_1} \ldots \phi^{i_n}
\ee{
    
The relevant vertex operators are\footnote{ Here the correlator is evaluated for generic $n\neq 0$ and the $n=0$ term is obtained by extrapolation. A direct evaluation of this term  is subtler since there are not enough insertions to completely fix the $SL(2,\R)$ world-sheet invariance.}  
\bea
W_{NS{-}NS}(z, \bar z) &=& c_{\rm NS} \,  (E R)_{MN}  \,e^{-\varphi} \psi^{M} e^{ik X} (z)  \, e^{-\varphi} \psi^{N} e^{ik  R X} (\bar z) \nn\\
V_{\xi(\phi)}({x}_a) &=& 
 \sum_{n=0}^\infty  \xi_{i_1\ldots i_n}  \,   \partial X^{i_1}  (x_1) \, 
\prod_{a=2}^n\, \int_{-\infty}^\infty \,    {d{x}_a\over 2\pi  }  \,   \partial X^{i_a}  ({x}_a)
\eea
with $E=h + b$  the polarization tensor containing the fluctuation $h$ of the metric and $b$ of the B-field and $c_{NS}$ a normalisation constant.    The momentum of the closed string state will be labeled by $k$.  Henceforth we denote by $X(z)$ the holomorphic part of the closed string field $X(z,\bar z)=X(z)+R X(\bar z)$ with $R$ the reflection matrix implementing the boundary conditions on the disk. More precisely, $R$ is a diagonal matrix  with plus one ({+}1) and minus one ({-}1) along the Neumann 
  and Dirichelet directions, respectively. In particular for a disk of type $0$, we have plus along $t,y_I$ and minus along $x_i,\tilde y_I$. Consequently $kR=-k$.  
   We choose length units whereby  $\alpha'=2$. 
  Moreover we exploit $SL(2,\mathbb{R})$ invariance of the disk to fix the positions of the closed string and the first of the open string insertions.
  Other open string vertices are integrated along the real line to take into account all possible orderings of the insertions.
  
  The resulting string amplitude can be written as 
\begin{equation}
\mathcal{A}_{NS{-}NS,\xi(\phi)}=  \langle c(z) c(\bar z) c(z_1) \rangle  \left\langle  W_{NS{-}NS}(z,\bar z)      V_{\xi(\phi)}   \right\rangle
\end{equation}
    The basic contributions to the correlators are 
  \bea
  \langle c(z) c(\bar z) c(z_1) \rangle &=&(z-\bar z)(\bar z-z_1)(z_1-z)\nn\\
  \left\langle   e^{\ii k X}(z)  e^{-\ii k  X}  ( \bar z)   \, \partial X^{i_a}  ({x}_a)  \right\rangle   &=&    \ii  k^{i_a}  \,    \left( {1\over z-{x}_a}-{1\over \bar z-{x}_a  } \right) =
   \ii k^{i_a}   \,   { \bar z-z\over |z-{x}_a|^2 }   \nn\\
  \left\langle   e^{-\varphi} \psi^{M} (z) e^{-\varphi} \psi^{N}  ( \bar z)     \right\rangle   &=&  { \delta^{MN} \over (z-\bar z)^2}    \label{nsns1}
  \eea
 Using 
  \bea
 \int_{-\infty}^\infty   {d{x}_a}  {   (\bar z-z) \over   | z-{x}_a|^2 } &=&  - 2 \pi \ii
  \eea
  one finds 
  \be
\begin{boxed}{\mathcal{A}_{NS{-}NS, \xi(\phi) } = \ii \, c_{\rm NS}\, \, {\rm tr} ({E}R)  {  \xi(  k)  } }\end{boxed}    
 \ee
  The asymptotic deviation from the flat metric can be extracted from   
  \be
  \delta \tilde g_{MN} (k) =  \left( -{  \ii \over k^2} \right) \sum_{n=0}^\infty  \,  \,  {\delta \mathcal{A}_{NS{-}NS, \phi^n } \over \delta h_{MN} } =  c_{\rm NS}
  { \xi(  k) \over  k^2 } \, (\eta R)_{MN}  
   \ee
    After Fourier transform one finds 
    \be
     \delta  g_{MN}= \int {d^3 k \over (2\pi)^3}  \delta\tilde g_{MN}=    -\ft12 (\eta R)_{MN}  \, \delta L (x)
    \ee
  with the identification 
    \be
     \delta L (x) = -2 \,c_{\rm NS} \, \int {d^3 k \over (2\pi)^3}   { \xi(  k) \over  k^2 }   e^{ikx}    
  \label{lfourier}
    \ee 
  We conclude that the function $\xi(\phi)$ codifies the complete multipole expansion of $L(x)$ in terms of  vacuum
   expectation values for untwisted scalar fields 
   \be
  \xi_{i_1\ldots   i_n}= \langle {\rm tr} \, \phi_{i_1} \ldots \phi_{i_n} \rangle 
  \ee
   In particular,  the harmonic function for a single D3-brane at position $a$, is  realized by taking
   \be
    \xi (\phi) \sim e^{\ii \, a\,  \phi}
   \ee
     that  reproduces the multipole expansion (\ref{1multi}) of $L(x)$ according to (\ref{lfourier}). 
   Similar results can be found by considering disks ending on D3-branes  associated to the harmonic functions $L_I$.
   
    We conclude by observing that D- and F-flatness conditions may restrict the choices for  $\xi_{i_1\ldots   i_n}$. Indeed for pointlike branes, the flatness conditions require $[\phi_i,\phi_j]=0$, so the matrices $\phi_i$ can be simultaneously diagonalized and written in terms 
   of $N_0$ positions in $\R^3$  \cite{Klebanov:1999tb}.

 \subsection{R{-}R amplitude}

Next we consider the R{-}R string amplitude
\begin{equation}
{\cal A}_{R{-}R, \xi(\phi)}=  \langle c(z) c(\bar z) c(z_1) \rangle  \left\langle  W_{R{-}R}(z,\bar z) 
 V_{\xi(\phi)}   \right\rangle
\end{equation}
with
\bea
W_{R{-}R}(z, \bar z) &=& c_{\rm R}\,   \mathcal{(FR)}_{\Lambda \Sigma} \, e^{-{\varphi\over 2} }  \, S^{\Lambda} e^{ik X}(z)\, e^{-{\varphi \over 2} } \, S^{\Sigma} \, e^{ik R X}(\bar z)  \nn\\
V_{\xi(\phi)}   &=&  \sum_{n=0}^\infty \, \xi_{i_1\ldots i_n}  e^{-\varphi} \psi^{i_1}  (z_1)    \prod_{a=2}^n  \int_{-\infty}^{\infty}   \, {d{x}_a \over 2\pi }  \, \partial X^{i_a}  ({x}_a)
\eea
Here 
\be
S^\Lambda=e^{{\ii \over 2} (\pm \varphi_1 \pm \varphi_2 \ldots \pm \varphi_5)}   \qquad {\rm with~even}~\#~{\rm of}~-'s 
\ee
represents a ten-dimensional spin field of positive chirality with $\Lambda = 1, \ldots 16$ and 
\be
{\cal F}=\sum_{n} \ft{1}{n!} F_{M_1\ldots M_n} \Gamma^{M_1\ldots M_n} \qquad ~~~~~~ {\cal R} =  \Gamma^{t y_1 y_2 y_3}
\ee

where $F$ are the R-R field strengths and ${\cal R} $ is the reflection matrix in the spinorial representation.

The basic contributions to the correlators are given by the first two lines of (\ref{nsns1}) and
  \bea
&&  \left\langle   e^{-{\varphi \over 2}}  \, S^{\Lambda} (z)  e^{-{\varphi \over 2}}  \, S^{\Sigma} \,   ( \bar z)     e^{-\varphi }\, \psi^M(z_1)   \right\rangle   =  {1\over \sqrt{2} } { (\Gamma^M)^{\Lambda \Sigma} \over (z-\bar z) |z-z_1|^2} 
  \eea
 Altogether  one finds
 \be
\begin{boxed}{\mathcal{A}_{R{-}R, \xi(\phi) } = \ii \,{c_{\rm R} \over \sqrt{2}}   \, {\rm tr}_{16} (  {\cal C  R} ) {   \xi( k)} ={16 \, \ii \, c_{\rm R} \over \sqrt{2}} \, { \xi( k)} 
C_{t y_1 y_2 y_3}}\end{boxed}
 \ee
  where we wrote ${\cal F}=\ii k_i \Gamma^i {\cal C}$. Here  and below we will  always restrict ourselves to the `electric' components $C_{t M_1 M_2 M_3 }$ since the remaining `magnetic' components are determined
  by the self-duality of $F_5$. The   R-R fields  at the disk level is then determined from the corresponding variation of the string amplitudes
  \be
  \delta \tilde C_{t y_1 y_2 y_3}=\left( -{  \ii \over k^2} \right) \sum_{n=0}^\infty \,  {\delta \mathcal{A}_{R{-}R, \xi(\phi)  } \over \delta C_{t y_1 y_2 y_3}  } = 8 \,\sqrt{2}     \,c_{\rm R} \, {\xi(k)  
   \over k^2 } 
   \ee
   Choosing 
   \be
   c_{\rm R} ={ c_{\rm NS} \over 4 \sqrt{2}}      \label{crcns}
   \ee
   and using (\ref{lfourier}) one finds
   \be
   \delta C_{t y_1 y_2 y_3}=- \delta L(x)
   \ee
    in agreement with (\ref{metricv})  for an arbitrary harmonic function $L(x)$ specified by $\xi(k) $ via   (\ref{lfourier}). 
 
 \section{ String amplitudes and 2-charge microstates  }
 
 Next we consider K solutions.  
 At  leading order in $\alpha_{D3}$, K solutions (\ref{d34k}) reduce to
\bea
 \delta g_{MN} dx^M dx^N & = & - 2\, dt \, w     - 2  \,  K \,  dy_3 \tilde dy_3 +\ldots  \label{deltag2} \\
  \delta C_4 &=&  (K\,  dt \wedge  dy_3    -w   \wedge  d\tilde y_3)\wedge  (dy_1 \wedge d\tilde y_2+ d\tilde y_1 \wedge d y_2  )   
\eea
  with 
  \be
  {*_3}dw=-dK
  \ee
  and $K$ a harmonic function starting at order $|x|^{-2}$.  For example one can take $K$  to be
\be
 K=    { v_i x_i \over  |x|^3}     + \dots  \label{k3}
\ee
 while \footnote{For instance, for $v_1=v_2=0$, $v_3\neq 0$ one has $w=  v_3  { (x_1 \, dx_2 - x_2 \, dx_1)/|x|^3} $.}
\be
w= \epsilon_{ijk} \,v_i  { x_j \, dx_k \over  |x|^3} + \dots  
\ee

\subsection{NS{-}NS amplitude}

 In this section we compute the disk amplitude generating the metric of the K solutions. K solutions will be associated to string fermionic bilinears 
 localised at the intersections of two D3-branes. The relevant string amplitude has been computed in \cite{Giusto:2009qq} for the case of D1-D5-branes.  
 Here we review this computation, adapting it to the case of D3-branes, and include the effect of untwisted open string insertions in order to account
 for higher multi-pole modes of the harmonic function $K$.   
 
  We consider a non-trivial open string condensate  
  \be
  {\cal O}^{AB}={ \rm tr } \, \bar \mu^{(A}  \mu^{B)} \, \xi(\phi)  
  \ee
    with $\bar \mu^A$ 
   a fermionic excitation of the open string  starting from D3$_0$ and ending on D3$_3$ and  $\mu^B$ a fermion excitation of the open string with opposite orientation. Here and below we denote with trace the sum over Chan-Paton indices along the boundary of the disk.
   We organise states in representations of the $SO(1,5)$ Lorentz symmetry rotating the six-dimensional hyper-plane along which the two stacks of branes are NN or DD, {\it i.e.}
    the space-time and the $y_3,\tilde y_3$ directions.  
 Upper (lower)  indices $A=1,\ldots 4$ and $M=1,\dots 6$  run over the right (left) spinor and vector representations of this group, respectively. 
    The projection onto the symmetric part is required by the irreducibility of the string diagram since the anti-symmetric part of the fermionic bilinear 
   produces a   scalar field $\phi_{[AB]}$ and consequently a  factorisation channel.  
 
The relevant disk amplitude can be written as
 \bea
\mathcal{A}^{NS{-}NS}_{\mu^2, \xi(\phi)} &=&   \int     dz_4  \langle c(z_1)\,c(z_2)\,c(z_3) \rangle  \left\langle V_{\bar \mu}(z_1) \,V_{\mu}(z_2)  \,   W(z_3,z_4)  \,   V_{ \xi(\phi)}  \right \rangle
\eea
with 
    \bea
 V_{\bar\mu}(z_1) &=& \bar \mu^{A} \,e^{-\varphi /2}\,  S_{A}   \sigma_2 \sigma_3  \nn  \\
 V_{\mu}(z_2) &=&  \mu^{B} \,e^{-\varphi /2}\,  S_{B}    \sigma_2 \sigma_3     \\
 V_{\xi(\phi)}    &=&     \sum_{n=0}^\infty \xi_{i_1\ldots i_n}  \prod_{a=1}^{n}\, \int_{-\infty}^\infty \, {dx_a \over 2\pi }  \,   \partial X^{i_a}  (x_a)      \nn\\
W(z_3,z_4) &=& c_{\rm NS}\, ({E}R)_{MN}  e^{-\varphi} \psi^{M} \, e^{\ii k X}(z_3)
(\partial  X^{N}+i \, k  \psi \, \psi^{N}) e^{-\ii k  X}(z_4) \nn
 \eea
Here $\sigma_I$ denotes the $\mathbb{Z}_2$-twist field along the $I^{\rm th}$ $T^2$ inside $T^6$ with conformal dimension 1/8 and
 \be
 S_{A}=e^{ \pm \frac{1}{2}(\ii \varphi_3 \pm\ii \varphi_4 \pm\ii \varphi_5) }        \qquad {\rm even~number~-'s} 
 \ee   
the spin field on the six-dimensional plane along which D3-branes are NN or DD.
 
 To evaluate the correlator one can consider a specific component and then use $SO(6)$ invariance to reconstruct its covariant form.  
For instance, if we take  $A=B=\ft12(+++)$, the open string condensate contributes a net charge $+2$ along the first three complex directions, so that only the $\psi{:}\psi \psi{:}$ term can contribute to the correlator. The relevant correlators are  
 \bea
 &&  dz_4\langle c(z_1)\,c(z_2)\,c(z_3)\rangle = dz_4\, z_{12}\, z_{23}\, z_{31}
 = dw\, (z_{14} z_{23})^2   \nn\\
 &&     \left\langle e^{ik X}(z_3) e^{ -ik  X}(z_4)  \,  V_{\xi(\phi)}  \right \rangle  =\xi(k) \nn\\
&& \left\langle   e^{-\varphi/2}(z_1)   e^{-\varphi/2}(z_2)   e^{-\varphi}(z_3)     \right\rangle =  z_{12}^{-1/4} (z_{13} z_{23}  )^{-1/2}  \nn\\  
&& \left\langle    \sigma_2 \sigma_3 (z_1)   \sigma_2 \sigma_3 (z_2)   \right\rangle =   z_{12}^{ -1/2  } \nn \\
&& \left \langle S_{A}(z_1)  S_{B}(z_2)  \psi^{ M} (z_3)  \psi^{N}\psi^{P}(z_4)  \right\rangle  =  
   {  \Gamma^{MNP}_{AB} \,  z_{12}^{3/4}        \over {2\sqrt{2}} (z_{13} z_{23} )^{1/2}   z_{14} z_{24}   }   
   \label{corr2ns}
 \eea
 with $w={z_{13} z_{24} \over z_{14} z_{23} }$. We notice that the scalar fields  factorize from the rest, since they are always oriented along the non compact space directions, while twist fields are along $T^6$.
 
  The $z_i$ dependence  boils down to the worldsheet integral
\be
{\cal I}=\int_\gamma  {dw \over w} =2\pi i   \label{ii2pi}
\ee
where $\gamma$ is a small contour around the origin.  On the other hand the open string condensate can be written as 
  \be
 \left\langle{\rm tr} \, \bar \mu^{(A}  \mu^{B)}  \right \rangle  = \frac{c_{\cal O} }{3!}\, v_{MNP}\, (\Gamma^{MNP})^{AB}  \label{mumu}
 \ee
 with $v_{MNP}$ a self-dual tensor  in six-dimensions   and $c_{\cal O}$ a normalisation constant. 
 Combining (\ref{mumu}) with (\ref{corr2ns}) and taking
 \be
 c_{\cal O} ={1\over \sqrt{2}\, {\cal I} \, c_{\rm NS}}    \label{co}
 \ee
  one finds
 \begin{equation}
\begin{boxed}{
\mathcal{A}^{NS{-}NS}_{\mu^2, \xi(\phi)}  =  \ft{1}{3!}\, (E R)_{MN} k_P  \, v^{MNP}\,  \xi(k) 
} \end{boxed}
  \label{nsamp2}
\end{equation}
 As before, the factor $\xi(k)$ comes from untwisted insertions and is associated  to higher multi-pole 
 modes of the underlying harmonic function, so  it is enough to match the leading term, {\it i.e.} we set $\xi(k)=1$. 
 Assuming  a $v_{IJK}$ of the form 
  \be
 v_{y_3 \tilde{y}_3  3} = - v_{12t}= 4 \pi \, v   \label{v3}
  \ee
 and using (\ref{nsamp2}) one finds
\be
\delta \tilde g_{2t}=- 4\pi v\,  k_1     \qquad  \delta \tilde g_{1t}= 4\pi v \, k_2     \qquad     \delta \tilde g_{y_3 \tilde y_3}= -4\pi v\,  k_3     
\ee
  After Fourier transform one finds
 \be
 \delta g_{2t} = -v\,  { x_1 \over  |x|^3}      \qquad  \delta g_{1t}= v { x_2 \over  |x|^3}     \qquad     \delta  g_{y_3 \tilde y_3}= -v\,  { x_3 \over   |x|^3}     
 \ee 
   in agreement with (\ref{deltag2}) for the choice (\ref{k3}) with $v_i=\delta_{i3} \,v$. We conclude that the harmonic function $K_3$ describes  a particular component of the  fermion bilinear  condensates  connecting two branes parallel along the $(y_3,\tilde y_3)$-plane. Similarly $K_{1,2}$ represents  fermion bilinear condensates connecting branes parallel along
the 1 and 2 planes (tori). 

 We conclude by noticing that  other choices for the 2-charge condensate give rise to solutions that do not belong to the `eight-ful' harmonic family (\ref{d34}). In particular, taking non-zero values of $v_{t y_3 \tilde{y}_3}$, $v_{t y_3 i}$ or $v_{t \tilde{y}_3 i}$ give rise to solutions with non-trivial  $b_{ij}$, $g_{{y}_3 t}$ or $b_{\tilde{y}_3 t}$ components, respectively. These components can be matched against a more general 2-charge solution obtained from \cite{Kanitscheider:2007wq} after two T-dualities. 
   We refer the reader to  \cite{Giusto:2009qq} for a detailed  match in the T-dual D1-D5 description.

\subsection{R{-}R amplitude}

Next we consider the disk amplitude with  the insertion of a R{-}R vertex operator
 \bea
\mathcal{A}^{R{-}R}_{\mu^2,\xi(\phi)} &=&    \int     dz_4  \langle c(z_1)\,c(z_2)\,c(z_3)\rangle   \left\langle V_{\bar\mu}(z_1) V_{\mu}(z_2)     W_{R{-}R}(z_3,z_4)     V_{\xi(\phi)} \right \rangle
\eea
with
\bea
W_{R{-}R}(z_3,z_4)&=&  c_{\rm R} \, \mathcal{(FR)}_{CD}   (e^{-{\varphi \over 2} }C^{C}C^{\dot{\alpha}} e^{\ii k X})(z_3)(e^{-{\varphi \over 2}} C^{D}C_{\dot{\alpha}}e^{- \ii k  X})(z_4)\nn
\eea

\bea
 C^{A} &=&e^{ \pm  \frac{1}{2}(\ii \varphi_3 \pm\ii \varphi_4 \pm\ii \varphi_5) }        \qquad {\rm odd~number~-'s}  \\
  C_{\dot \alpha}&=&e^{ \pm {\ii\over 2}(\varphi_4-\varphi_5) }  
   \eea  
The  ghost and untwisted bosonic correlators are given again by the first two lines in (\ref{corr2ns}) while the fermionic and twist-field correlators are
\bea
&& \langle S_{(A}(z_1) S_{B)}(z_2)  C^{C}(z_3) C^{D}(z_4)  \rangle = \delta_{(A}^{C} \delta_{B)}^{D} \left( \frac{z_{12} z_{34}}{z_{13} z_{14}z_{23} z_{24}} \right)^{3/4}   \nn\\
&&  \langle \sigma_2 \sigma_3(z_1) \sigma_2 \sigma_3(z_2) \rangle=z_{12}^{-1/2} \nn\\
 &&  \langle \prod_{i=1}^4 e^{-\varphi/2}(z_i)  \rangle=\prod_{i<j}^4  z_{ij}^{-1/4}   \nn\\
 && \langle C^{\dot{\alpha}}(z_3) C_{\dot{\alpha}}(z_4) \rangle=2 \, z_{34}^{-1/2}
\eea
 Again one finds that the amplitude is proportional to ${\cal I}$ given in (\ref{ii2pi}) and  can be written in the form  
\begin{equation}
\begin{boxed}{
\mathcal{A}^{R{-}R}_{\mu^2,\xi(\phi)} ={ 1 \over \, 3! \, 4}  \, v_{MNP} \, {\rm tr}_4 \left( \mathcal{ FR}  \Gamma^{MNP} \right)  \xi(k)
} \end{boxed}
\label{rrd3d3}
\end{equation}
  where   we  used  (\ref{mumu}), (\ref{co}) and (\ref{crcns}).  
  Taking $\xi(k)=1$, specializing to the condensate in (\ref{v3})  and focussing on the $C_{t M_1M_2 M_3}$ components one finds
 \be
\mathcal{A}^{R{-}R}_{\mu^2,\xi(\phi)} =    4\pi v  \left(   F_{ t 3  y_1 \tilde y_2  y_3 } +  F_{ t 3  \tilde y_1  y_2  y_3 }  \right) 
\ee
Varying with respect to the four-form potential one finds
   \be
   \delta \tilde C_{t y_1 \tilde y_2 y_3 } =   \delta \tilde C_{t \tilde y_1 y_2 y_3 }=   k_3 \,  4\pi v   
   \ee
 reproducing the Fourier transform of  (\ref{deltag2})  for the choice (\ref{k3})  with $v_i=\delta_{i3} \,v$.  

\section{String amplitudes and 4-charge microstates  }

Finally we consider M solutions.   At  leading order in $\alpha_{D3}$, M solutions (\ref{d34mu}) reduce to
\bea
&&  \delta g_{MN} dx^M dx^N = 2 M \,  \left( dy_1\, \tilde dy_1+dy_3\, \tilde dy_3  \right)  +\ldots \nn  \\
&& \delta C_4 =-  M\, dt  \wedge\, (     dy_1\wedge d \tilde y_2  \wedge d y_3  +   d\tilde y_1\wedge d  \tilde y_2\wedge d\tilde y_3 )  \nn\\
&& ~~~~~~~~ + w_2 \wedge ( dy_1   \wedge dy_2 \wedge dy_3 +     d\tilde y_1 \wedge dy_2\wedge  d\tilde y_3   )  
   +\ldots\label{solm}
\eea
  with  $M$ a harmonic function starting at order $|x|^{-3}$.   For example one can take $M$  to be of the form
\be
 M=    v_{ij} {   3 \,x_i\, x_j -\delta_{ij} |x|^2 \over   |x|^5}     + \dots
\ee

\subsection{NS{-}NS amplitude}

We consider now string amplitudes on a disk with boundary on all four types of D3-branes. In particular we consider the insertions of four fermions $\mu_a$  
starting on a D3-brane of type $(a)$ and ending on a D3-branes of type $(a+1)$ with $a=0,1,2, 3$ (mod 4),  in a cyclic order.  
We notice that unlike the case of two boundaries now the condensate is complex. Indeed, even if each intersection preserves ${\cal N}=2$ SUSY (1/4 BPS), so that each fermion $\mu_a$ comes together with its charge conjugate $\bar \mu_a$, the overall configuration preserves only ${\cal N}=1$ SUSY (1/8 BPS) and fermions connecting all four type of D3 brane form two pairs of opposite chirality. The charge conjugate condensate can be defined by replacing each $\mu_a$ with its charge conjugate field $\bar \mu_a$  and running along the boundary of the disk  with the same cyclic order but in  reversed sense.  The real and imaginary parts of the string amplitude can be selected by turning on the
real and imaginary parts of the condensate respectively. 

We consider the following NS{-}NS amplitude 
\bea
\mathcal{A}^{NS{-}NS}_{\mu^4, \xi(\phi)}&=&  \langle c(z_1)\,c(z_2)\,c(z_4)\rangle \int dz_3 dz_5 dz_6  \\
&& \left \langle V_{\mu_1}(z_1) V_{\mu_2}(z_2) V_{\mu_3}(z_3) V_{\mu_4}(z_4)    W_{NSNS}(z_5,z_6)  V_{\xi(\phi)} \right\rangle\nn
\eea
with 
\bea
 V_{\mu_1}(z_1) &=& \mu_1^{\alpha} \,e^{-\varphi /2}\,  S_{\alpha}  S_{1}  \,\sigma_{2} \sigma_3 (z_1)   \\
 V_{\mu_2}(z_2) &=& \mu_2^{\beta} \,e^{-\varphi /2}\,  S_{\beta}  S_{3} \sigma_{1} \sigma_2 (z_2)  \nn\\
 V_{\mu_3}(z_3) &=& \mu_3^{\dot\alpha} \,e^{-\varphi /2}\,  C_{\dot\alpha} \bar S_{1} \sigma_{2}\sigma_3   (z_3)  \nn\\
 V_{\mu_4}(z_4) &=& \mu_4^{\dot\beta} \,e^{-\varphi /2}\,  C_{\dot\beta} \bar S_{3} \sigma_{1} \sigma_2 (z_4)   \nn\\
  V_{\xi(\phi)}  &=&     \sum_{n=0}^\infty \xi_{i_1\ldots i_n}  \prod_{a=1}^{n}\, \int_{-\infty}^\infty \, {dx_a \over 2\pi }  \,   \partial X^{i_a}  (x_a)     \nn\\
W_{NSNS}(z_5,z_6) &=& c_{\rm NS} (E R)_{MN}  (\partial X^{M}-\ii k \cdot \psi \psi^{M})e^{\ii kX}(z_5)(\partial  X^{N}+\ii k  \cdot  \psi \psi^{N}) e^{-\ii k  X}(z_6) \nn
 \eea
 We use the notation
 \be
 S_I=e^{\ii \varphi_I \over 2} \qquad \bar S_I=e^{-{\ii \varphi_I \over 2} } \qquad   S_{\alpha}=e^{ \pm {\ii\over 2}(\varphi_4 +\varphi_5 ) }   \qquad   C_{\dot \alpha}=e^{ \pm {\ii\over 2}(\varphi_4-\varphi_5) }    
 \ee   
for internal and spacetime spin fields. The condensate is now the tensor
  \be
 {\cal O}^{\alpha\beta\dot\alpha\dot\beta} =  {\rm tr}  \, \mu_1^{(\alpha} \, \mu_2^{\beta)}\,  \bar\mu_3^{(\dot\alpha}\,  \bar\mu_4^{\dot\beta)} \, \xi(\phi)\label{omcond}
  \ee
where  the sum over all the Chan-Paton indices $\mu^1_{i_1}{}^{i_2} \mu^2_{i_2}{}^{i_3} \bar\mu^3_{i_3}{}^{i_4} \bar\mu^4_{i_4}{}^{i_1}$ is understood. As before  we discard components proportional to $\epsilon^{\alpha\beta}$ and 
  $\epsilon^{\dot\alpha\dot\beta}$ to ensure the irreducibility of the string diagram. As a result $ {\cal O}^{\alpha\beta\dot\alpha\dot\beta}$ transforms in the $({\bf 3},{\bf 3})$ of the
  $SU(2)_L\times SU(2)_R$ Lorentz group and can be better viewed as   a symmetric and traceless tensor $v^{\mu\nu}$. We notice that even though the presence of the four different D3-branes breaks the Lorentz symmetry down
  to $SO(4)$, the string amplitude is invariant under $SO(4)\times U(1)^3$ since the  $ U(1)^3$ rotations of the three internal tori do not change the relative angles between branes. We notice also that space-time chirality determines the internal $U(1)^3$ charges of the twisted fermions. 
     In particular $\mu_1^{\alpha}$ and  $\bar\mu_3^{\dot \alpha}$ have opposite charges with respect to the first $U(1)$ while
     $\mu_2^{\alpha}$ and  $\bar\mu_4^{\dot \alpha}$  have opposite charges with respect to the third internal $U(1)$.  
  Therefore the open string condensate $v_{\mu \nu}$ carries no $U(1)^3$ charges.   
        
To evaluate the correlator we specialize to $\alpha=\beta=\ft12(++)$ and  $\dot{\alpha}=\dot{\beta}=\ft12 (-+)$. Only the four fermion piece of the closed string vertex can 
    compensate for the net $+2$ charge  of the open string condensate and two of these fermions have to be taken along  the fifth plane.    
        In addition since the net internal $U(1)^3$ charge of the condensate is zero, the only choices for the   $\psi^M \psi^N$ fermions in the closed string vertex 
  are $(M,N)=(I,\bar I)$ with $I=1,2,3,4$  

For example, taking  $(M,N)=(1,\bar 1)$ the relevant correlators are   
 \bea
 && \langle c(z_1)\,c(z_2)\,c(z_4)\rangle =  z_{12}\, z_{24}\, z_{41}  \nn \\
  &&     \left\langle e^{ik X}(z_3) e^{ -ik  X}(z_4)  \,  V_{\xi(\phi)}  \right \rangle  =\xi(k) \nn\\
&& 
\left\langle   \sigma_{3} (z_1)   \sigma_{1} (z_2)   \sigma_{3} (z_3)   \sigma_{1}  (z_4)  \right\rangle 
= (z_{13} z_{24})^{-1/4 }    \nn\\
&& \left\langle \prod_{j=1}^4 e^{-\varphi/2}(z_j) \right\rangle =
\prod_{i<j}  z_{ij}^{-1/4 } \nn\\ 
&& 
\left\langle  S_1   (z_1)   \bar S_1   (z_3)  \psi^1 (z_5)     \psi^{\bar 1} (z_6) \right\rangle \left\langle  S_3  (z_2)   \bar S_3  (z_4)      \right\rangle 
 =\ft{1}{2}  z_{13}^{-{1\over 4} }  z_{56}^{-1}   \left( { z_{15} z_{36} \over z_{16} z_{35}   } \right)^{1/2}   ( z_{24})^{-1/4 }   \nn\\
 && \left\langle    \sigma_{2} (z_1) \sigma_{2} (z_2)   \sigma_{2} (z_3)  \sigma_{2}  (z_4)  \right\rangle =   f\left(\ft{z_{14} z_{23} }{z_{13} z_{24} } \right)  
 \left({ z_{13} z_{24} \over  z_{12} z_{23} z_{34} z_{41} } \right)^{1/4 }  \nn\\
&& \left \langle S_{(\alpha}(z_1) S_{\beta)}(z_2) C_{(\dot \alpha}(z_3) C_{\dot\beta)}(z_4) \psi^{ \mu} (z_5)     \psi^{ \nu} (z_6)  \right\rangle   =  
  { (z_{12} z_{34} )^{1/2} z_{56} \over  
2 \prod_{i=1}^4 (z_{i5} z_{i6})^{1/2}  }  \sigma^{(\mu}_{\alpha \dot{\alpha}  }  \sigma^{\nu)}_{\beta \dot \beta}   \label{corr4ns}
 \eea
 with $f(x)$ the four twist correlator \cite{Cvetic:2003ch, Bianchi:2007rb,Anastasopoulos:2011hj} 
 \be
 f(x)= { \Lambda(x)   \over   (F(x)F(1-x))^{1/2} } 
 \ee
 where $F(x) ={}_2F_1(1/2,1/2; 1; x)$ is a hypergeometric function\footnote{Equivalently, $F(x) = \vartheta_2(it(x))/\vartheta_3(it(x))$ with $t(x) = F(1-x)/F(x)$. } and 
 \be
 \Lambda (x) =  \sum_{n_1,n_2}  e^{- {2\pi \over \alpha'}   \left[   { F(1-x) \over F(x) } \, n_1^2 R_1^2  +    { F(x) \over F(1-x) } \, n_2^2 R_2^2   \right]}    
 \ee
 accounts for the classical contribution  associated to world-sheet instantons\footnote{ 
 In the supergravity limit $\sqrt{\alpha'} << R_{1,2}$, worldsheet instantons are exponentially suppressed  and one can simply take $\Lambda(x)\to 1$}. Assembling all pieces together and taking
  \be
  z_1=-\infty \qquad z_2=0\qquad z_3=x \qquad z_4=1 \qquad z_5=z \qquad z_6=\bar{z}
  \ee
  one finds that the string amplitude is proportional to the integral 
      \bea
 I_1 &=&  \int_0^1    \,     dx  \,  f(x)\,  {\cal I}_1  (x)
   \label{ii0}
  \eea
  with
  \be
   {\cal I}_1 (x) = \int_{\mathbb{C}^+}  { d^2z \over |z(1-\bar z)|(x- z)} 
 \label{i1}
  \ee
  Similarly for other choices of $(M,N)=(I,\bar I)$, $I=2,3,4$,  one finds integrals of the form (\ref{ii0}) with ${\cal I}_1$ replaced by 
  \bea
  {\cal I}_2 (x)&=&   \int_{\mathbb{C}^+}  { d^2z \over |z(1-\bar z)(x- z)|} \nn\\
  {\cal I}_3 (x) &=& \int_{\mathbb{C}^+}  { d^2z \over \bar z (1- z) |x- z|}  \nn\\ 
  {\cal I}_4 (x)  &=&   \int_{\mathbb{C}^+}  { d^2z \over \bar z(1-z)(x- z)} 
  \label{i234}
 \eea
 The integrals (\ref{ii0}),  (\ref{i1}), (\ref{i234}) can be computed numerically for arbitrary values of the radii and one always finds  a finite result,  with ${\cal I}_{2}, {\cal I}_4$ real and   ${\cal I}_{1}, {\cal I}_3$
 purely imaginary\footnote{Similar integrals appear in \cite{Stieberger:2009hq}.}. Moreover the following relation holds
 \be
  {\cal I}_{1}= {\cal I}_3
 \ee 
   Noticing that real and imaginary parts of the amplitude contribute to symmetric $({E}R)_{(MN)}$ and antisymmetric parts $({E}R)_{[MN]}$ 
 we conclude that a purely imaginary  $ v^{ij} $ generates the string amplitude 
 \be
\begin{boxed}{
\mathcal{A}^{NS{-}NS}_{\mu^4, \xi(\phi)}=   \left[ ({E}R)_{[1\bar 1]} +  ({E}R)_{[3\bar 3]}  \right] k_{i} k_{j}  \,    v^{ij} \, \xi(k)
}\end{boxed}  
\label{a4nsns}
\ee 
with
 \be
 \left\langle {\rm tr} \, \mu_1^{(\alpha} \, \mu_2^{\beta)}\,  \bar\mu_3^{(\dot\alpha}\,  \bar\mu_4^{\dot\beta)}  
     \right\rangle = {2 \pi v^{ij}  \over c_{\rm NS} \, {\cal I}_1} \,  \sigma_i^{\alpha\dot\alpha } 
   \bar\sigma_j^{\beta \dot\beta}   
  \ee
   We notice that $({E}R)_{[1\bar 1]}$ and $({E}R)_{[3\bar 3]} $ always involve a Neumann and a Dirichlet direction, so only the metric contributes
   to the antisymmetric part of the matrix.  Varying with respect to the symmetric part $h_{ij}$  of  the polarization tensor $E=h+b$ one finds
\bea
\delta \tilde g_{1\bar 1 } &=&\delta \tilde g_{3\bar 3 }  = -2\pi   \ii \,  v^{ij} \, \frac{\, k_{i} k_{j} }{k^2}    \, \xi(k) 
\eea
 that reproduces (\ref{solm}) after Fourier transform. 
 
We conclude this section by noticing that there  other choices for the 4-charge condensate besides (\ref{omcond}) lead to solutions that go beyond the eight harmonic function family 
(\ref{d34}). On the one hand one can consider  different orderings of the branes along the disk. These condensates lead again to solutions of type M but with internal coordinates that are permuted among one another. Notice that some of these solutions can look very different from each other since $L_I$ and $V$ do not enter in a symmetric fashion.
On the other hand, turning on the real part of the condensate (\ref{omcond}) will produce solutions with a non-trivial B-field given in terms of the real worldsheet integrals ${\cal I}_2$ and ${\cal I}_4$.  Finally turning one can consider condensates of type\footnote{The open string condensate $\check{\cal O}^{(\alpha\beta\gamma\delta)}$ does not couple to massless closed string states.}
 \bea
 {\cal \widetilde O}^{\alpha\dot\alpha\beta\dot\beta}&=&{\rm tr}  \, \mu_1^{(\alpha} \,  \bar\mu_2^{(\dot\alpha}\, \mu_3^{\beta)}\,  \bar\mu_4^{\dot\beta)} \,  \nn \\
\hat{\cal O}^{(\alpha\beta\gamma)\dot\beta} &=&{\rm tr}  \, \mu_1^{(\alpha} \,  \mu_2^{\beta}\, \mu_3^{\gamma)}\,  \bar\mu_4^{\dot\beta} 
 \eea
These condensates generate a new class of solutions with off-diagonal terms mixing either two different $T^2$'s inside $T^6$ or space-time and internal components. We leave it as an open question to establish  whether or not an extended family  of supergravity solution accounting for these brane excitations can be explicitly written.

\subsection{R{-}R amplitude}

Finally we consider the amplitude with the insertion of a R{-}R vertex operator
\bea
\mathcal{A}^{R{-}R}_{\mu^4, \xi(\phi)}&=&  \langle c(z_1)\,c(z_2)\,c(z_4)\rangle \int dz_3 dz_5 dz_6  \\
&& \left \langle V_{\mu_1}(z_1) V_{\mu_2}(z_2) V_{\mu_3}(z_3) V_{\mu_4}(z_4)    W_{R{-}R}(z_5,z_6)  V_{\xi(\phi)} \right\rangle\nn
\label{ampl4D3R{-}R}
\eea
with 
\bea
 W_{R{-}R}^{(-1/2,+1/2)}(z_5,z_6)&=& c_{\rm R}\,  (\mathcal{FR} \Gamma_M)_\Lambda{}^\Sigma \, e^{-{\varphi \over 2} } S^{\Lambda}\, e^{ik X} (z_5) e^{\varphi \over 2}   (\partial X^{M}+ i  k  \psi \psi^{M} )  C_{\Sigma} \, e^{ -ik  X}  (z_6)\nn
\eea
the R-R vertex operator in the asymmetric   (-1/2,+1/2) super-ghost picture\footnote{Similarly to the `canonical'  (-1/2,-1/2) picture, thanks to the extra $\Gamma^M$, $W_{R{-}R}^{(-1/2,+1/2)}$ can be expressed in terms of field-strengths $F_n$ solely, unlike in the (-3/2,-1/2) picture that involves the potentials $C_{n{-}1}$ \cite{Bianchi:1991eu}.}.
  Notice that $ \Psi^M_\Sigma (k) = {:}k\psi \psi^{M}C_{\Sigma}{:}$ is a world-sheet primary field of dimension 13/8, belonging to the {\bf 128} of $SO(1,9)$, with $C_{\Sigma}$ the spin field of negative chirality.
  Again taking $\alpha=\beta=\ft12(++)$, $\dot\alpha=\dot\beta=\ft12(-+)$, it is easy to see that the net charge $-2$ of the condensate can be only compensated by the  
   $ k  \psi \psi^{M}$ term in the closed string vertex   with   $S^{\Lambda}$ and $C_\Sigma$ both carrying charge $-\ft12$ along the fifth 
   direction. This leads to eight choices for  $S^{\Lambda}$. 
   On the other hand, once $S^{\Lambda}$ is chosen, the charges of $ \Psi^M_\Sigma (k) = {:}k\psi \psi^{M}C_{\Sigma}{:}$ are completely determined by charge conservation.  
   For example, for the choice
  \bea
  S^\Lambda &=&  e^{{\ii\over 2} (\varphi_1+\varphi_2-\varphi_3+\varphi_4-\varphi_5)}  \nn\\
   k \psi \, \psi^M\, C_\Sigma &=&k_{\bar 5} \psi^{\bar 5} \psi^{\bar 1} e^{{\ii\over 2} (\varphi_1-\varphi_2+\varphi_3-\varphi_4-\varphi_5)} 
  \eea
  one finds again the first four lines in (\ref{corr4ns}), while the last  four lines are replaced by
   \bea
&& \langle  \prod_{i=1}^5 e^{-{\varphi \over 2}}(z_i) e^{\varphi \over 2}(z_6)  \rangle=\prod_{j=1}^5  z_{j6}^{1\over 4} \prod_{i<j}^5  z_{ij}^{-{1\over 4} }  \\ 
 && \left\langle  S_1   (z_1) S_3 (z_2) \bar S_1 (z_3) \bar S_3   (z_4)   S_1 S_2  \bar S_3 (z_5)\, \bar S_1 \bar S_2  S_3     (z_6) \right\rangle =   \left( { z_{15} z_{45} z_{26} z_{36}  \over z_{13} z_{24} z_{25} z_{35} z_{16} z_{46} z_{56}^3  }\right)^{1\over 4}  \nn\\
 && \left \langle S_{(\alpha}(z_1) S_{\beta)}(z_2)    C_{(\dot \alpha}(z_3) C_{\dot\beta)}(z_4)   C_{\dot \gamma} (z_5)   \psi^\mu S_{ \gamma} (z_6)  \right\rangle \nn\\
 &&~~~~~~~= \ft{ 1 }{\sqrt{2}} 
     \left( {z_{12} z_{34}  z_{56} \over z_{35} z_{45}   z_{36} z_{46}  z_{16}^2 z_{26}^2  } \right)^{1\over 2}  \epsilon_{\gamma(\alpha} \sigma_{\beta)(\dot\beta}^\mu   \epsilon_{\dot \alpha)\dot\gamma }\nn
   \eea
Combining the various correlators and taking
  \be
  z_1=-\infty \qquad z_2=0\qquad z_3=x \qquad z_4=1 \qquad z_5=z \qquad z_6=\bar{z}
  \ee
 one finds that the string amplitude is again given by the integral (\ref{ii0}) involving ${\cal I}_1$.  A similar analysis can be performed for other choices of spin-field components (labelled by
 $\Lambda, \Sigma$) leading to similar answers in terms of the four characteristic integrals ${\cal I}_1, \ldots {\cal I}_4$ on $\C$. 
 The results are summarised in table 2.   Since ${\cal I}_2$, ${\cal I}_4$ 
 are real and ${\cal I}_1, {\cal I}_3$ purely imaginary,  a purely imaginary condensate selects the ${\cal I}_1$, ${\cal I}_3$ and ${\cal \bar I}_1$, ${\cal \bar I}_3$ components.   
 \begin{table}[htdp]
\begin{center}
\begin{tabular}{|c|c|c|c|c|}
\hline
 & ${\cal I}_1 $ & ${\cal I}_2$ & ${\cal I}_3$ & ${\cal I}_4$ \\ 
 \hline
 $\Lambda$ &  $\ft12(++-+-) $  & $ \ft12(+++--)$ & $ \ft12(-+++-)$ & $\ft12(+-++-)$ \\
 \hline
 & ${ \cal \overline{I} }_1 $ & ${\cal \bar I}_2$ & ${\cal \bar I}_3$ & ${\cal \bar I}_4$ \\ 
 \hline
  $\Lambda$    &  $\ft12(--+--)$  & $ \ft12(---+-)$ & $ \ft12(+----)$ & $\ft12(-+---)$ \\
 \hline
\end{tabular}
\end{center}
\label{tspinor}
\caption{Contributions to the string correlator of the various spinor components.  }
\end{table}%
  The resulting string amplitude can then be written as
 \be
\begin{boxed}
{ \mathcal{A}^{RR}_{\mu^4, \xi(\phi)}=\ft{ 1}{4 } {\rm tr}_{16} (\mathcal{F\, R} \, {\cal P}\, \Gamma^i    )  2 \pi  v_{ij}  k_j \, \xi(k)
}\end{boxed}
 \ee
 with 
 \be
 {\cal P}   = \ft12(1-\Gamma^{y_1\tilde y_1 y_3\tilde y_3} )  \Gamma^{y_2 \tilde y_2} 
 \ee
 a projector on  ${\cal I}_1$, ${\cal I}_3$ and ${\cal \bar I}_1$, ${\cal \bar I}_3$ components   
  with $+\ii$ and $-\ii$ eigenvalues respectively. These are precisely the eigenvalues of the matrix ${\cal P}$ justifying our claim. 
 Using ${\cal R}=\Gamma^{t y_1 y_2 y_3}$ one   finds
 \be
\begin{boxed} { \mathcal{A}^{RR}_{\mu^4, \xi(\phi)}=2 \pi \ii  \,  k_i v_{ij}  k_j \, \xi(k)  \left(  C_{t  y_1  \tilde y_2  y_3} + C_{ t   \tilde y_1 \tilde y_2 \tilde y_3}          \right) }\end{boxed}
 \ee
 Taking derivatives with respect to $C_4$ one finds agreement with (\ref{solm}) after Fourier transform.


\section{Conclusions and outlook}

We have provided a direct link between open string condensates and a large class of micro-state geometries associated to four-dimensional BPS black holes consisting of D3-branes intersecting on $T^6$. For specific choices of the condensates,  we match the order $g_s$  of the micro-state solutions (\ref{d34})  against disk amplitudes involving a NS-NS or R-R closed string state and zero, two or four twisted open string fermions besides an arbitrary number of untwisted scalars. Each mixed disk is associated to a harmonic function that for the class of supergravity solutions under consideration here can be  written as a linear combination of eight harmonic functions $H_a = \{L_I,V,K_I,M\}$. The eight functions describe the distribution of single, 2- and 4-intersections in the $\R^3$ non compact space directions.  
 
The M class of micro-states is particularly interesting  since the  number of disks with four different boundaries grows with the product of the four charges $N=Q_1\, Q_2\, Q_3 \,Q_4$. In a first approximation, the function $M$ can then be viewed as an $N$-center harmonic function with moduli space  spanned by the $N$ positions on $\R^3$  up to $S_N$-permutations.  One may then expect that he number of micro-states in this class  grow as $e^{2\pi \sqrt{N/6}}$,  the number of conjugacy classes of $S_N$ or equivalently the partitions of $N$,   at large N.  
  This suggests that the 4-charge micro-states identified in this work may account for a large number of states contributing to the (putative) black hole entropy. A quantitative study of this issue remains one of the most challenging and exciting questions left open by our work. 

Although we have worked mostly with D3-branes that are either parallel (along some directions) or perpendicular (along other directions), our analysis can easily be adapted to branes intersecting at arbitrary angles and preserving  the same amount of supersymmetry (1/8 of the original 32 super-charges) \cite{Bertolini:1998mt,Bertolini:1999je} or branes intersecting in orbifolds \cite{Bertolini:2000ei,Bertolini:2000yaa}. The open strings living at the intersections will be described in terms of twist-fields \cite{Pesando:2014owa,Pesando:2012cx,Pesando:2011ce} that generalise the ${\bf Z}_2$ twist fields that appeared in the present analysis. We have explicitly checked that one-, two- and four- open string insertions source for classical supergravity profiles similar to those obtained in this paper. We defer a detailed analysis to future work. For the time being we would like to stress that off-diagonal components of the internal metric can mimic the effect of tilting the branes thus changing their intersection angles. Furthermore, it is amusing to observe that the 5 harmonic functions that appear in the static solutions found in \cite{Bertolini:1998mt,Bertolini:1999je, Bertolini:2000ei,Bertolini:2000yaa} can be identified with $V,L_I$ and $M$. This is consistent with the fact that setting $K_I=0$ produces micro-state geometries with no angular momentum.  
 
Our solution generating technique have useful physical applications, both for understanding the physics of multi-center extremal solutions and for constructing new smooth micro-states geometries for under-rotating extremal black holes in four and  higher dimensions.  Indeed, in addition to reproducing all known supergravity solutions in the `eightful' harmonic family for certain choices of the condensates, we have shown that other choices are possible that can generate new 4-charge micro-state geometries beyond the ones available in the literature.  

\section*{Acknowledgements}

We are particular grateful to S.~Giusto, S.~Mathur  and R~Russo for illuminating discussions and comments on the manuscript. We would  also like to thank A.~Addazi, M.~Bertolini, R.~Emparan, F.~Fucito, A.~Guerrieri, G.~Horowitz, O.~Lunin, L.~Martucci,  I.~Pesando,  Ya.~Stanev,  M.~Trigiante, C.-K.~Wen for interesting discussions.

 \begin{appendix}

\section{Conventions}

The basic bosonic and ghost correlators are ($\alpha'=2$)
\bea
\langle  X^{\mu}(z,\bar{z}) X^{\nu}(w,\bar{w}) \rangle   &=&  - \eta^{\mu \nu} \log |z-w|^2  \nn\\
\langle  \partial X^M(x)  e^{  i k_L X_L}(z) e^{  i k_R X_R}( \bar z)  \rangle &=&   { \ii\, k^M_L \over z-x}+ {   \ii\, k_R^M\over \bar z -x}   \nn\\
\langle     e^{  q_1 \varphi }(z) e^{   q_2 \varphi }(w)  \rangle  &=&  (z-w)^{-q_1 q_2}   \nn\\
\langle c(z) c(w)  \rangle  &=& (z-w) 
\eea
Fermions are bosonized according to
 \be
   \Psi^I=e^{i \varphi_I}={1\over \sqrt{2}} \,(\psi_{2I-1}+i \psi_{2I})   \quad \Rightarrow \quad    \Psi_M \,S^\Lambda\sim  \ft{1}{\sqrt{2z} } \, (\Gamma_M)^{\Lambda\Sigma} \, C_\Sigma
\ee
 and fermionic correlators can be computed with the help of
\bea
\langle     e^{ i q_1 \varphi_I}(z) e^{  i q_2 \varphi_J}(w)  \rangle  &=&  \delta_{IJ} (z-w)^{ q_1 q_2}   \nn\\
  \langle  \psi^{M}(z) \psi^{N}(w) \rangle &=&   \frac{ \eta^{\mu \nu} }{(z-w)} 
  \eea

  \section{ The supergravity solution} 
 
 \subsection{D0-D2-D4-D6 solution }
 
 A family of under-rotating multi-center black hole solutions corresponding to a general system of D0-D2-D4-D6 branes in type IIA theory wrapped on $T^6$ was constructed 
in  \cite{Balasubramanian:2006gi, Dall'Agata:2010dy}. The solutions are  parametrised by eight harmonic functions ($a=1,\ldots 8$, $I=1,2,3$)
\be
H_a= \{ V,K_I,L_I,M\}   
\ee
It is convenient to combine these functions into 
    \bea
 P_I & =&  {K_I \over V}  \nn\\
  Z_I  &=&  L_I +{ C_{IJK}\over 2} {K_J K_K\over V}  \nn\\
\mu  &=&  {M\over 2} +{L_I K_I \over 2\, V}+{ C_{IJK} \over 6} \, {K_I K_J K_K \over V^2}
  \eea
In term of these functions the  metric in the string frame and the various p-form potentials are written as  \cite{Dall'Agata:2010dy}
\bea
 ds^2_{10} &=&  -   e^{2U}( dt+w)^2 +e^{-2U} \,   \sum_{i=1}^3 dx_i^2    +     \sum_{I=1}^3  \frac{e^{-2U} }{V Z_I} ( dy_{I}^2+d\tilde y_{I}^2)  \nn\\
  e^{-2 \phi} &=& e^{6U} V^3 Z_1 Z_2 Z_3    \qquad  ~~~~~~~~~~~B_2 =  \sum_{I=1}^3    b_I  \,dy_I \wedge d\tilde y_I    \nn\\
     C_1 & =& \alpha_0 \quad\quad       ~~~~~~~~~~~~~~~~~~~~~~~~~\,  C_7 =\beta_0 \, \prod_{I=1}^3 dy_I \wedge d\tilde y_I   \label{giusto} \\
 C_3  &=&  \sum_{I=1}^3  \alpha_I   \wedge  \,dy_I \wedge d\tilde y_I     \qquad ~~~~~  C_5 ={|\epsilon_{IJK} | \over 2} \, \beta_I \wedge dy_J \wedge d\tilde y_J 
 \wedge dy_K \wedge d\tilde y_K  \nn 
  \eea  
 with
 \bea
  e^{-4U} &=&  Z_1 Z_2 Z_3 V-\mu^2  V^2 \qquad ~~~~~~~~~~ b_I =   \left( P_I-\frac{\mu}{Z_I} \right) \nn\\
 \alpha_0 &=& A-\mu\, V^2\, e^{4U}\, (dt+w) \qquad ~~~~
\alpha_I =  \left[  -\frac{(dt+w)}{Z_I} + b_I\, A+w_I   \right] \nn\\
\beta_0&=& -v_0 +{e^{-4U}\over V^2 Z_1 Z_2 Z_3}(dt+w)-b_I \,v_I+ b_1 \,b_2\, b_3 \, A +{|\epsilon_{IJK}|\over 2}\,b_I \,b_J\, w_K  \nn\\
\beta_I &=& -v_I + {|\epsilon_{IJK}|\over 2}\, \left(  {\mu\over Z_J Z_K}\, (dt+w)+b_J \, b_K\, A+2 b_J\, w_K  \right) 
 \eea
    and
  \bea
 {*_3}dA &=& d V     \qquad {*_3}dw_I = -d (K_I)  \qquad    {*_3}dv_0 =dM  \qquad    {*_3}dv_I =dL_I\nn\\
   \qquad  {*_3}dw  &=&  V d \mu-\mu dV-V Z_I dP_I  
  \eea
 Here $A$, $w$, $w_I$, $v_0$, $v_I$, $\alpha_0$ and $\alpha_I$ are one-forms in the (flat) 3-dimensional $x$-space.   The R-R field strengths are defined by
    \be
{F}_n = dC_{n-1} - H_3 \wedge C_{n-3} ,
\ee
 and satisfy the duality relations
 \be
 {*_{10}}F_2= F_8 \qquad       {*_{10}}F_4=- F_6
 \ee  
   We stress that the inclusion of $C_5$ and $C_7$ is required  in order to get a self-dual five-form field $F_5=dC_4$ after T-dualities. Only after the inclusion of these terms, the Buscher's rules map solutions of type IIA to solutions of IIB and viceversa. 
   
  \subsection{D3$^4$-solution}
 
The D0-D2-D4-D6 system can be mapped to a system of solely D3-branes applying  three T-dualities  along $\tilde y_1$, $\tilde y_2$ and $\tilde y_3$. 
 T-duality  mixes the supergravity fields according to the generalised Buscher rules~\cite{Lunin:2001fv}:
 \be
\begin{aligned}
g'_{zz} &= \frac{1}{g_{zz}}, \quad\quad\quad e^{2 \phi '} = \frac{e^{2 \phi}}{g_{zz}}, \quad\quad\quad g'_{z m} = \frac{B_{z m}}{g_{zz}}, \quad\quad\quad B'_{z m}= \frac{g_{z m}}{g_{zz}} \\[8pt]
g'_{m n} &= g_{m n}-\frac{g_{m z}\,g_{n z} - B_{m z}\,B_{n z}}{g_{zz}}, \quad B'_{m n}= B_{m n}- \frac{B_{m z}\,g_{n z}-g_{m z}\,B_{n z}}{g_{zz}}\\[8pt]
C'^{(n)}_{m_1 ... m_{n}} &= C^{(n+1)}_{m_1 ... m_{n} z} - n\,C^{(n-1)}_{[m_1 ... m_{n-1}}\,B_{m_n ] z} - n(n-1)\frac{C^{(n-1)}_{[m_1 ... m_{n-2}| z}\,B_{|m_{n-1}| z}\,g_{|m_{n} ] z}}{g_{zz}}\\[8pt]
 C'^{(n)}_{m_1 ... m_{n-1}  z} &=C^{(n-1)}_{m_1 ... m_{n-1}} - \left( n - 1 \right) \frac{C^{(n-1)}_{[ m_1 ... m_{n-2} | z}\,g_{z | m_{n-1} ]}}{g_{zz}} \label{tdual}
\end{aligned}
\ee
where by $z$ we denote the direction along with T-duality is performed and the metric is  understood in the string frame. 
  Applying this rules one finds that after T-dualities along $\tilde y_1$,  $\tilde y_2$ and  $\tilde y_3$ the dilaton and the B-field becomes trivial while the metric
  maps to
 \bea
 ds^2 &=& -   e^{2U}( dt+w)^2 +e^{-2U} \,  \sum_{i=1}^3 dx_i^2    +      \sum_{I=1}^3   \left[  { d y_I^2 \over  V e^{2U} Z_I }  + V e^{2U} Z_I  \,  \tilde e_I^2  \right] \label{metriciia}
 \eea
 with 
   \be
\tilde e_I=   d \tilde y_I   -  \left( P_I-\frac{\mu}{Z_I} \right)  d y_I    
   \ee
 Now let us consider the RR fields.
It is easy to see that the two terms involving the metric $g$ in the 
T-duality  transformations of RR fields  in  (\ref{tdual}) do not contribute to the metric (\ref{metriciia}). More precisely, the term proportional to  $C^{(n-1)}_{[m_1 ... m_{n-2}| z}\,B_{|m_{n-1}| z}$ cancels 
because $C$ and $B$ always share two indices while  the term proportional to $C^{(n-1)}_{[ m_1 ... m_{n-2} | z}\,g_{z | m_{n-1} ]}$  cancels because  the original metric is always diagonal along the two-torus along one of whose two coordinates the following T-duality transformation is performed.  
As a result, the T-duality transformations can be written in the simple form
\be
C^{'(n)}=C^{(n-1)}_{\perp} \wedge \tilde e_I +  C_{||}^{(n+1)} 
\ee
 where we decompose the R-R forms 
\be
C^{(n)}=C_{||}^{(n-1)} \wedge d\tilde y_I + C_{\perp}^{n} 
\ee
 into a components containing or not the direction $d\tilde y_I$ along which the T-duality is performed. After T-dualities along $\tilde y_1$,  $\tilde y_2$ and  $\tilde y_3$, one finds that  all RR fields are mapped to the four-form $C_4$ that can be written in the form
      \bea
 && C_4 = \alpha_0  \wedge \tilde e_1\wedge \tilde e_2\wedge \tilde e_3  +\ft12 \epsilon_{IJK}\,  \alpha_I  \wedge  dy_I \wedge \tilde e_J   \wedge \tilde e_K \nn\\
&&~~~~  + \beta_0  \wedge dy_1\wedge  dy_2\wedge  dy_3  +\ft12 \epsilon_{IJK}\,  \beta_I  \wedge  \tilde e_I \wedge dy_J   \wedge dy_K    \label{c4gen}
  \eea   
 One can check that the five form flux
 \be
 F_5=d\, C_4
 \ee
 obtained from  (\ref{c4gen}) is self-dual as required.   
\end{appendix}



\begin{thebibliography}{10}

\bibitem{Mathur:2009hf}
S.~D. Mathur, \emph{{The Information paradox: A Pedagogical introduction}},
  \href{http://dx.doi.org/10.1088/0264-9381/26/22/224001}{Class. Quant. Grav.
  {\bf 26} (2009)  224001},
\href{http://arxiv.org/abs/0909.1038}{{\tt arXiv:0909.1038 [hep-th]}}.

\bibitem{Mathur:2005zp}
S.~D. Mathur, \emph{{The Fuzzball proposal for black holes: An Elementary
  review}}, \href{http://dx.doi.org/10.1002/prop.200410203}{Fortsch. Phys. {\bf
  53} (2005)  793--827},
\href{http://arxiv.org/abs/hep-th/0502050}{{\tt arXiv:hep-th/0502050
  [hep-th]}}.

\bibitem{Sen:1995in}
A.~Sen, \emph{{Extremal black holes and elementary string states}},
  \href{http://dx.doi.org/10.1142/S0217732395002234}{Mod. Phys. Lett. {\bf A10}
  (1995)  2081--2094},
\href{http://arxiv.org/abs/hep-th/9504147}{{\tt arXiv:hep-th/9504147
  [hep-th]}}.

\bibitem{Strominger:1996sh}
A.~Strominger and C.~Vafa, \emph{{Microscopic origin of the Bekenstein-Hawking
  entropy}}, \href{http://dx.doi.org/10.1016/0370-2693(96)00345-0}{Phys. Lett.
  {\bf B379} (1996)  99--104},
\href{http://arxiv.org/abs/hep-th/9601029}{{\tt arXiv:hep-th/9601029
  [hep-th]}}.

\bibitem{Lunin:2001fv}
O.~Lunin and S.~D. Mathur, \emph{{Metric of the multiply wound rotating
  string}}, \href{http://dx.doi.org/10.1016/S0550-3213(01)00321-2}{Nucl. Phys.
  {\bf B610} (2001)  49--76},
\href{http://arxiv.org/abs/hep-th/0105136}{{\tt arXiv:hep-th/0105136
  [hep-th]}}.

\bibitem{Lunin:2002bj}
O.~Lunin, S.~D. Mathur, and A.~Saxena, \emph{{What is the gravity dual of a
  chiral primary?}},
  \href{http://dx.doi.org/10.1016/S0550-3213(03)00081-6}{Nucl. Phys. {\bf B655}
  (2003)  185--217},
\href{http://arxiv.org/abs/hep-th/0211292}{{\tt arXiv:hep-th/0211292
  [hep-th]}}.

\bibitem{Kanitscheider:2007wq}
I.~Kanitscheider, K.~Skenderis, and M.~Taylor, \emph{{Fuzzballs with internal
  excitations}}, \href{http://dx.doi.org/10.1088/1126-6708/2007/06/056}{JHEP
  {\bf 06} (2007)  056},
\href{http://arxiv.org/abs/0704.0690}{{\tt arXiv:0704.0690 [hep-th]}}.

\bibitem{Giusto:2009qq}
S.~Giusto, J.~F. Morales, and R.~Russo, \emph{{D1D5 microstate geometries from
  string amplitudes}}, \href{http://dx.doi.org/10.1007/JHEP03(2010)130}{JHEP
  {\bf 03} (2010)  130},
\href{http://arxiv.org/abs/0912.2270}{{\tt arXiv:0912.2270 [hep-th]}}.

\bibitem{Bena:2007kg}
I.~Bena and N.~P. Warner, \emph{{Black holes, black rings and their
  microstates}}, \href{http://dx.doi.org/10.1007/978-3-540-79523-0_1}{Lect.
  Notes Phys. {\bf 755} (2008)  1--92},
\href{http://arxiv.org/abs/hep-th/0701216}{{\tt arXiv:hep-th/0701216
  [hep-th]}}.

\bibitem{Skenderis:2008qn}
K.~Skenderis and M.~Taylor, \emph{{The fuzzball proposal for black holes}},
  \href{http://dx.doi.org/10.1016/j.physrep.2008.08.001}{Phys. Rept. {\bf 467}
  (2008)  117--171},
\href{http://arxiv.org/abs/0804.0552}{{\tt arXiv:0804.0552 [hep-th]}}.

\bibitem{Mathur:2008nj}
S.~D. Mathur, \emph{{Fuzzballs and the information paradox: A Summary and
  conjectures}},
\href{http://arxiv.org/abs/0810.4525}{{\tt arXiv:0810.4525 [hep-th]}}.

\bibitem{Giusto:2011fy}
S.~Giusto, R.~Russo, and D.~Turton, \emph{{New D1-D5-P geometries from string
  amplitudes}}, \href{http://dx.doi.org/10.1007/JHEP11(2011)062}{JHEP {\bf 11}
  (2011)  062},
\href{http://arxiv.org/abs/1108.6331}{{\tt arXiv:1108.6331 [hep-th]}}.

\bibitem{Giusto:2012yz}
S.~Giusto, O.~Lunin, S.~D. Mathur, and D.~Turton, \emph{{D1-D5-P microstates at
  the cap}}, \href{http://dx.doi.org/10.1007/JHEP02(2013)050}{JHEP {\bf 02}
  (2013)  050},
\href{http://arxiv.org/abs/1211.0306}{{\tt arXiv:1211.0306 [hep-th]}}.

\bibitem{Bena:2011uw}
I.~Bena, J.~de~Boer, M.~Shigemori, and N.~P. Warner, \emph{{Double, Double
  Supertube Bubble}}, \href{http://dx.doi.org/10.1007/JHEP10(2011)116}{JHEP
  {\bf 10} (2011)  116},
\href{http://arxiv.org/abs/1107.2650}{{\tt arXiv:1107.2650 [hep-th]}}.

\bibitem{Giusto:2012jx}
S.~Giusto and R.~Russo, \emph{{Perturbative superstrata}},
  \href{http://dx.doi.org/10.1016/j.nuclphysb.2012.12.012}{Nucl. Phys. {\bf
  B869} (2013)  164--188},
\href{http://arxiv.org/abs/1211.1957}{{\tt arXiv:1211.1957 [hep-th]}}.

\bibitem{Giusto:2013bda}
S.~Giusto and R.~Russo, \emph{{Superdescendants of the D1D5 CFT and their dual
  3-charge geometries}}, \href{http://dx.doi.org/10.1007/JHEP03(2014)007}{JHEP
  {\bf 03} (2014)  007},
\href{http://arxiv.org/abs/1311.5536}{{\tt arXiv:1311.5536 [hep-th]}}.

\bibitem{Bena:2014qxa}
I.~Bena, M.~Shigemori, and N.~P. Warner, \emph{{Black-Hole Entropy from
  Supergravity Superstrata States}},
  \href{http://dx.doi.org/10.1007/JHEP10(2014)140}{JHEP {\bf 10} (2014)  140},
\href{http://arxiv.org/abs/1406.4506}{{\tt arXiv:1406.4506 [hep-th]}}.

\bibitem{Bena:2015bea}
I.~Bena, S.~Giusto, R.~Russo, M.~Shigemori, and N.~P. Warner, \emph{{Habemus
  Superstratum! A constructive proof of the existence of superstrata}},
  \href{http://dx.doi.org/10.1007/JHEP05(2015)110}{JHEP {\bf 05} (2015)  110},
\href{http://arxiv.org/abs/1503.01463}{{\tt arXiv:1503.01463 [hep-th]}}.

\bibitem{Denef:2000nb}
F.~Denef, \emph{{Supergravity flows and D-brane stability}},
  \href{http://dx.doi.org/10.1088/1126-6708/2000/08/050}{JHEP {\bf 08} (2000)
  050},
\href{http://arxiv.org/abs/hep-th/0005049}{{\tt arXiv:hep-th/0005049
  [hep-th]}}.

\bibitem{Denef:2002ru}
F.~Denef, \emph{{Quantum quivers and Hall / hole halos}},
  \href{http://dx.doi.org/10.1088/1126-6708/2002/10/023}{JHEP {\bf 10} (2002)
  023},
\href{http://arxiv.org/abs/hep-th/0206072}{{\tt arXiv:hep-th/0206072
  [hep-th]}}.

\bibitem{Bates:2003vx}
B.~Bates and F.~Denef, \emph{{Exact solutions for supersymmetric stationary
  black hole composites}},
  \href{http://dx.doi.org/10.1007/JHEP11(2011)127}{JHEP {\bf 11} (2011)  127},
\href{http://arxiv.org/abs/hep-th/0304094}{{\tt arXiv:hep-th/0304094
  [hep-th]}}.

\bibitem{Balasubramanian:2006gi}
V.~Balasubramanian, E.~G. Gimon, and T.~S. Levi, \emph{{Four Dimensional Black
  Hole Microstates: From D-branes to Spacetime Foam}},
  \href{http://dx.doi.org/10.1088/1126-6708/2008/01/056}{JHEP {\bf 01} (2008)
  056},
\href{http://arxiv.org/abs/hep-th/0606118}{{\tt arXiv:hep-th/0606118
  [hep-th]}}.

\bibitem{deBoer:2008fk}
J.~de~Boer, F.~Denef, S.~El-Showk, I.~Messamah, and D.~Van~den Bleeken,
  \emph{{Black hole bound states in AdS(3) x S**2}},
  \href{http://dx.doi.org/10.1088/1126-6708/2008/11/050}{JHEP {\bf 11} (2008)
  050},
\href{http://arxiv.org/abs/0802.2257}{{\tt arXiv:0802.2257 [hep-th]}}.

\bibitem{Dall'Agata:2010dy}
G.~Dall'Agata, S.~Giusto, and C.~Ruef, \emph{{U-duality and non-BPS
  solutions}}, \href{http://dx.doi.org/10.1007/JHEP02(2011)074}{JHEP {\bf 02}
  (2011)  074},
\href{http://arxiv.org/abs/1012.4803}{{\tt arXiv:1012.4803 [hep-th]}}.

\bibitem{Lunin:2015hma}
O.~Lunin, \emph{{Bubbling geometries for AdS$_{2}$Ã S$^{2}$}},
  \href{http://dx.doi.org/10.1007/JHEP10(2015)167}{JHEP {\bf 10} (2015)  167},
\href{http://arxiv.org/abs/1507.06670}{{\tt arXiv:1507.06670 [hep-th]}}.

\bibitem{Rychkov:2005ji}
V.~S. Rychkov, \emph{{D1-D5 black hole microstate counting from supergravity}},
  \href{http://dx.doi.org/10.1088/1126-6708/2006/01/063}{JHEP {\bf 01} (2006)
  063},
\href{http://arxiv.org/abs/hep-th/0512053}{{\tt arXiv:hep-th/0512053
  [hep-th]}}.

\bibitem{Krishnan:2015vha}
C.~Krishnan and A.~Raju, \emph{{A Note on D1-D5 Entropy and Geometric
  Quantization}}, \href{http://dx.doi.org/10.1007/JHEP06(2015)054}{JHEP {\bf
  06} (2015)  054},
\href{http://arxiv.org/abs/1504.04330}{{\tt arXiv:1504.04330 [hep-th]}}.

\bibitem{Lunin:2002iz}
O.~Lunin, J.~M. Maldacena, and L.~Maoz, \emph{{Gravity solutions for the D1-D5
  system with angular momentum}},
\href{http://arxiv.org/abs/hep-th/0212210}{{\tt arXiv:hep-th/0212210
  [hep-th]}}.

\bibitem{Klebanov:1999tb}
I.~R. Klebanov and E.~Witten, \emph{{AdS / CFT correspondence and symmetry
  breaking}}, \href{http://dx.doi.org/10.1016/S0550-3213(99)00387-9}{Nucl.
  Phys. {\bf B556} (1999)  89--114},
\href{http://arxiv.org/abs/hep-th/9905104}{{\tt arXiv:hep-th/9905104
  [hep-th]}}.

\bibitem{Cvetic:2003ch}
M.~Cvetic and I.~Papadimitriou, \emph{{Conformal field theory couplings for
  intersecting D-branes on orientifolds}},
  \href{http://dx.doi.org/10.1103/PhysRevD.70.029903,
  10.1103/PhysRevD.68.046001}{Phys. Rev. {\bf D68} (2003)  046001},
  \href{http://arxiv.org/abs/hep-th/0303083}{{\tt arXiv:hep-th/0303083
  [hep-th]}}.
[Erratum: Phys. Rev.D70,029903(2004)].

\bibitem{Bianchi:2007rb}
M.~Bianchi and J.~F. Morales, \emph{{Unoriented D-brane Instantons vs Heterotic
  worldsheet Instantons}},
  \href{http://dx.doi.org/10.1088/1126-6708/2008/02/073}{JHEP {\bf 02} (2008)
  073},
\href{http://arxiv.org/abs/0712.1895}{{\tt arXiv:0712.1895 [hep-th]}}.

\bibitem{Anastasopoulos:2011hj}
P.~Anastasopoulos, M.~Bianchi, and R.~Richter, \emph{{Light stringy states}},
  \href{http://dx.doi.org/10.1007/JHEP03(2012)068}{JHEP {\bf 03} (2012)  068},
\href{http://arxiv.org/abs/1110.5424}{{\tt arXiv:1110.5424 [hep-th]}}.

\bibitem{Stieberger:2009hq}
S.~Stieberger, \emph{{Open \& Closed vs. Pure Open String Disk Amplitudes}},
\href{http://arxiv.org/abs/0907.2211}{{\tt arXiv:0907.2211 [hep-th]}}.

\bibitem{Bianchi:1991eu}
M.~Bianchi, G.~Pradisi, and A.~Sagnotti, \emph{{Toroidal compactification and
  symmetry breaking in open string theories}},
\href{http://dx.doi.org/10.1016/0550-3213(92)90129-Y}{Nucl. Phys. {\bf B376}
  (1992)  365--386}.

\bibitem{Bertolini:1998mt}
M.~Bertolini, M.~Trigiante, and P.~Fre, \emph{{N=8 BPS black holes preserving
  1/8 supersymmetry}},
  \href{http://dx.doi.org/10.1088/0264-9381/16/5/305}{Class. Quant. Grav. {\bf
  16} (1999)  1519--1543},
\href{http://arxiv.org/abs/hep-th/9811251}{{\tt arXiv:hep-th/9811251
  [hep-th]}}.

\bibitem{Bertolini:1999je}
M.~Bertolini, P.~Fre, and M.~Trigiante, \emph{{The Generating solution of
  regular N=8 BPS black holes}},
  \href{http://dx.doi.org/10.1088/0264-9381/16/9/315}{Class. Quant. Grav. {\bf
  16} (1999)  2987--3004},
\href{http://arxiv.org/abs/hep-th/9905143}{{\tt arXiv:hep-th/9905143
  [hep-th]}}.

\bibitem{Bertolini:2000ei}
M.~Bertolini and M.~Trigiante, \emph{{Regular BPS black holes: Macroscopic and
  microscopic description of the generating solution}},
  \href{http://dx.doi.org/10.1016/S0550-3213(00)00216-9}{Nucl. Phys. {\bf B582}
  (2000)  393--406},
\href{http://arxiv.org/abs/hep-th/0002191}{{\tt arXiv:hep-th/0002191
  [hep-th]}}.

\bibitem{Bertolini:2000yaa}
M.~Bertolini and M.~Trigiante, \emph{{Microscopic entropy of the most general
  four-dimensional BPS black hole}},
  \href{http://dx.doi.org/10.1088/1126-6708/2000/10/002}{JHEP {\bf 10} (2000)
  002},
\href{http://arxiv.org/abs/hep-th/0008201}{{\tt arXiv:hep-th/0008201
  [hep-th]}}.

\bibitem{Pesando:2014owa}
I.~Pesando, \emph{{Correlators of arbitrary untwisted operators and excited
  twist operators for $N$ branes at angles}},
  \href{http://dx.doi.org/10.1016/j.nuclphysb.2014.06.010}{Nucl. Phys. {\bf
  B886} (2014)  243--287},
\href{http://arxiv.org/abs/1401.6797}{{\tt arXiv:1401.6797 [hep-th]}}.

\bibitem{Pesando:2012cx}
I.~Pesando, \emph{{Green functions and twist correlators for $N$ branes at
  angles}}, \href{http://dx.doi.org/10.1016/j.nuclphysb.2012.08.016}{Nucl.
  Phys. {\bf B866} (2013)  87--123},
\href{http://arxiv.org/abs/1206.1431}{{\tt arXiv:1206.1431 [hep-th]}}.

\bibitem{Pesando:2011ce}
I.~Pesando, \emph{{The generating function of amplitudes with $N$ twisted and L
  untwisted states}}, \href{http://dx.doi.org/10.1142/S0217751X15501213}{Int.
  J. Mod. Phys. {\bf A30} (2015) no.~21, 1550121},
\href{http://arxiv.org/abs/1107.5525}{{\tt arXiv:1107.5525 [hep-th]}}.

\end{thebibliography}

\providecommand{\href}[2]{#2}\begingroup\raggedright\endgroup

\end{document}